%% file: main.tex
\documentclass[fleqn,usenatbib]{mnras}

\usepackage{newtxtext,newtxmath}

\usepackage[T1]{fontenc}

\usepackage{graphicx}	
\usepackage{amsmath}	
\usepackage{xspace}
\usepackage{xcolor}
\usepackage{enumerate}

\newcommand{\vect}[1]{\boldsymbol{#1}} 

\newcommand{\refeqstyle}[1]{(#1)\xspace}
\newcommand{\refeqlabel}{equation}
\newcommand{\Refeqlabel}{Equation}
\newcommand{\reffiglabel}{Fig.~}
\newcommand{\reftablabel}{Table~}
\newcommand{\refseclabel}{\S~}
\newcommand{\refapplabel}{Appendix}
\newcommand{\refeq}[1]{\refeqlabel~\refeqstyle{\ref{#1}}} 

\newcommand{\refeqp}[1]{\refeqlabel~\ref{#1}\xspace} 
\newcommand{\reffig}[1]{\reffiglabel\ref{#1}\xspace} 
\newcommand{\reftab}[1]{\reftablabel\ref{#1}\xspace} 
\newcommand{\refsec}[1]{\refseclabel\ref{#1}\xspace} 

\newcommand{\starsmasher}{\texttt{StarSmasher}\xspace}
\newcommand{\mesa}{\texttt{MESA}\xspace}

\newcommand{\modelone}{{\xspace}{\it R30N1}\xspace}
\newcommand{\modeltwo}{{\xspace}{\it R37N1}\xspace}
\newcommand{\modelthree}{{\xspace}{\it R37N2}\xspace}
\newcommand{\modelfour}{{\xspace}{\it R37N3}\xspace}
\newcommand{\modelfive}{{\xspace}{\it R10N1}\xspace}

\title[Stellar models]{Simulating a stellar contact binary merger -- I. Stellar models}

\author[R.\ W.M.\ Hatfull, N.\ Ivanova and J.\ C.\ Lombardi Jr.]{
Roger W.M. Hatfull,$^{1}$\thanks{E-mail: \href{mailto:rogerhatfull@gmail.com}{rogerhatfull@gmail.com}}
Natalia Ivanova,$^{1}$
James C.\ Lombardi Jr.$^{2}$
\\
$^{1}$University of Alberta, Edmonton, AB, T6G 2R3, Canada\\
$^{2}$Allegheny College, Meadville, PA, 16335, United States
}

\date{Accepted XXX. Received YYY; in original form ZZZ}

\pubyear{2021}

\begin{document}
\label{firstpage}
\pagerange{\pageref{firstpage}--\pageref{lastpage}}
\maketitle

\input{Abstract.tex}

\input{Section1.tex}
\input{Section2.tex}
\input{Section3.tex}

\input{Section4.tex}

\input{Section5.tex}
\section*{Data availability statement}
The data underlying this article will be shared on reasonable request to the corresponding author.

\input{Acknowledgements.tex}

\bibliographystyle{mnras}
\bibliography{references} 

\bsp	
\label{lastpage}
\end{document}

%% file: Abstract.tex
\begin{abstract}
We study the initial conditions of a common envelope (CE) event resulting in a stellar merger.
A merger's dynamics could be understood through its light curve, but no synthetic light curve has yet been created for the full evolution.
Using the Smoothed Particle Hydrodynamics (SPH) code \texttt{StarSmasher}, we have created three-dimensional (3D) models of a 1.52\,$M_\odot$ star that is a plausible donor in the V1309~Sco progenitor.
The integrated total energy profiles of our 3D models match their initial one-dimensional (1D) models to within a 0.1\,per~cent difference in the top 0.1\,$M_\odot$ of their envelopes.
We have introduced a new method for obtaining radiative flux by linking intrinsically optically thick SPH particles to a single stellar envelope solution from a set of unique solutions.
For the first time, we calculated our 3D models' effective temperatures to within a few per~cent of the initial 1D models, and found a corresponding improvement in luminosity by a factor of $\gtrsim10^6$ compared to ray tracing.
We let our highest resolution 3D model undergo Roche-lobe overflow with a 0.16\,$M_\odot$ point-mass accretor ($P\simeq1.6$\,days) and found a bolometric magnitude variability amplitude of $\sim0.3$ -- comparable to that of the V1309~Sco progenitor.
Our 3D models are, in the top 0.1\,$M_\odot$ of the envelope and in terms of total energy, the most accurate models so far of the V1309~Sco donor star.
A dynamical simulation that uses the initial conditions we presented in this paper can be used to create the first ever synthetic CE evolution light curve.
\end{abstract}

\begin{keywords}
hydrodynamics
-- methods: numerical
-- radiative transfer
-- stars: low-mass
-- binaries: close
\end{keywords}

%% file: Section1.tex
\section{Introduction}%
Luminous Red Novae (LRNe) are transient events that are identified by a rise in luminosity to within the nova-supernova gap $10^{39}$--$10^{41}$\,ergs\,s$^{-1}$ \citep{2012PASA...29..482K}, a distinct red color that becomes redder with time, and a lengthy luminosity plateau after the main outburst, which is sometimes followed by a secondary maximum.
Approximately a few LRNe occur every 10 years in the Galaxy \citep{2014MNRAS.443.1319K,2020MNRAS.492.3229H}.
A list of LRNe observations is provided in \citet{2020MNRAS.492.3229H}, with several more observed since: AT 2018bwo \citep{2021arXiv210205662B}, AT 2019zhd \citep{2021A&A...646A.119P}, and AT 2020hat and AT 2020kog \citep{2020arXiv201110590P}.
\citet{2003ApJ...582L.105S} argued that, for outburst events similar to V838~Mon, the energetics may be explained by a merger of two stars.
The best evidence that followed was the observed binary orbital period decay of V1309~Sco prior to its outburst \citep{2011A&A...528A.114T}.
While the V1309~Sco outburst was significantly less luminous than the V838~Mon outburst, they share some specific features with several other observed transient events: color, presence of a plateau, peak luminosity, plateau duration dependence, photosphere temporal evolution, rapid decline, and differences between spectral and apparent photosphere expansion velocities.
It has been shown that the specific features of this new class of LRNe can be explained within the model of wavefront of cooling and recombination in the ejecta due to a CE event \citep{2013Sci...339..433I}, uniting the observed transient events into one class of physical events.

CE evolution \citep{1976IAUS...73...75P} is a phase of evolution of a close binary when a shared, partially non co-rotating gas envelope, is formed that surrounds the spiraling-in binary.
The drag forces dissipate the binary's orbital energy into the CE, shrinking the distance between the companion star and the core of the donor.
The outcome is either a complete merger to a single coalesced star or, if the CE is successfully ejected, a more compact binary.
CE evolution is an important transformational stage that is considered to be responsible for the formation of many X-ray binaries \citep{1976IAUS...73...35V}, binary pulsars \citep{1976ApJ...207..574S}, cataclysmic variables, Type Ia supernovae \citep{1993PASP..105.1373I,2021A&A...645A..54K}, hot subdwarfs \citep{2002MNRAS.336..449H,2020A&A...642A.180P}, and some planetary and pre-planetary nebulae \citep{2000ASPC..199..201B,2001ApJ...557..256S,2012ApJ...746..100S,2017NatAs...1E.117J,2018A&A...617A.129K,2021A&A...646A...1K}.
CE evolution is also one possible explanation for the existence of blue and red stragglers \citep{1964MNRAS.128..147M,2019MNRAS.486.1220F,2019A&A...624A.128B}, and may lead to the formation of progenitors of gravitational wave sources \citep{1993MNRAS.260..675T,2003MNRAS.342.1169V,2012ApJ...759...52D,2016Natur.534..512B,2017NatCo...814906S,2018MNRAS.481.4009V}.

The fact that CE events have been observed in action has manifested a new step in development of common envelope physics, as for the first time the theories of how it proceeds can be verified by observations.
However, although we now have a plethora of observations compared to 10 years ago, a direct comparison has not been made yet.
The problem lies in producing the observational signature from numerical simulations, a synthetic light curve, and matching it to an observed light curve.
Radiative flux calculations coupled with three-dimensional (3D) hydrodynamics have mostly been done under stratified gas temperature distribution assumptions and have been limited in scope to the pre-outburst phase
\citep{2016MNRAS.461.2527P,2017ApJ...850...59P,2017ApJS..229...36G,2017MNRAS.471.3200M,2021arXiv210205662B}.
Simplified stratified gas temperature provides `observed' temperatures that are too hot \citep[see, e.g.][]{2017ApJS..229...36G}.
To properly resolve the gas temperature distribution, it is required to use a prohibitively large spatial resolution (as many as $\sim10^{10}$ elements; see \refsec{sec:single_stars} for further discussion).
The only other phase of CE evolution for which some progress has been made is the plateau \citep{2017MNRAS.470.2339L}.

The far-reaching goal of our study is to reconstruct light curves of stellar mergers and CE events in close binaries.
The focus of this paper is on the creation of the initial stellar models that will later be merged with a secondary star in future work.
In \refsec{sec:sph_stellar_models} we describe the simulation code we use and discuss our method for creating stellar models whose total energy profiles match with an error not exceeding 0.1\,per~cent that of their initial one-dimensional (1D) parent stars in the upper 0.1\,$M_\odot$ of their envelopes.
In \refsec{sec:calculating_flux_in_SPH} we identify the difficulties in using ray tracing when calculating the radiative flux from our simulations and address these challenges with our new `envelope fitting' method. 
In \refsec{sec:results} we present our results for the luminosities of our 3D stellar models, for which we observe an improvement by a factor of $10^4$ in using envelope fitting over ray tracing.
In the same section we also obtain the luminosity of the V1309~Sco progenitor binary at time of contact and quantify our uncertainty.
    

%% file: Section2.tex
\section{Stellar Models}\label{sec:sph_stellar_models}%
To model our 3D stars we use the Smoothed Particle Hydrodynamics (SPH) code \starsmasher\footnote{\starsmasher is available at https://jalombar.github.io/starsmasher/.} \citep{2010MNRAS.402..105G}.
We first evolve a 1D star using the 1D detailed stellar evolution code \mesa \citep{Paxton2011, Paxton2013, Paxton2015, Paxton2018, Paxton2019}. 
Our 3D simulations are initialized using the 1D profiles taken at several evolutionary points (see \refsec{sec:initialization}).
For each star, a family of initialized 3D models are relaxed until they approach hydrostatic equilibrium, \refsec{sec:relaxation}.
Then we select the relaxed models that best match their 1D progenitor energy profiles, \refsec{sec:matching_the_mesa_model}. 
The properties of the best matched 3D models are discussed in \refsec{sec:relaxed_models}.

\subsection{Stellar model initialization}\label{sec:initialization}%
We consider a star with mass $M=1.52$\,$M_\odot$ as the donor in a plausible V1309~Sco progenitor binary system $1.52+0.16$\,$M_\odot$, from {\reftablabel}1 of \citet{2011A&A...531A..18S}.
We evolve the donor star using \mesa version 9793, with solar chemical composition $Z=0.02$ and $X=0.7$ until it leaves the main sequence and enters the base of the red giant branch\footnote{The inlists are available at the \mesa marketplace.}.

We build our base models, which we name as \modelone, \modeltwo, and \modelfive from three points in the $M=1.52$\,$M_\odot$ star's evolution after it has developed a convective outer envelope and expanded to radii $R_\text{1D}=3.055$\,$R_\odot$, $3.715$\,$R_\odot$, and $10.00$\,$R_\odot$ respectively.
\modeltwo fits a plausible Roche-lobe-overflowing V1309~Sco progenitor for the observed pre-merger orbital period \citep{2011A&A...531A..18S}.

We construct our 3D models from our base models using $N\simeq1\times10^5$ particles.
We create two additional 3D models, \modelthree and \modelfour, using the same base model as \modeltwo but with $N\simeq2\times10^5$ and $N\simeq3\times10^5$ correspondingly.
To facilitate the best match between 1D and 3D stellar models, we use tabulated equations of state (TEOS) with \starsmasher, as described in \citet{2015MNRAS.450L..39N}.
The TEOS is built upon the \mesa module for EOS, which uses a blend of the OPAL \citep{Rogers2002}, SCVH \citep{Saumon1995}, PTEH \citep{Pols1995}, HELM \citep{Timmes2000}, and PC \citep{Potekhin2010} EOSs.

We initialize our 3D models with particles in a hexagonal close-packed (hcp hereafter) lattice \citep{1976itss.book.....K}, a configuration that is stable to perturbations \citep{2006ApJ...640..441L}.
We assign particle masses
\begin{equation}
    m_i = \frac{4}{3}\pi R_\text{1D}^3 \frac{\rho(r_i)}{N}\ ,
    \label{eq:m_i}
\end{equation}
where $\rho(r_i)$ is the density from the 1D model interpolated at each particle $i$'s distance from the center $r_i$ using a cubic spline interpolator.
We also assign initial particle specific internal energies $u_i$ and mean molecular weights $\mu_i$ by the same interpolation method.

By choosing particles of unequal masses, we allow for a better mass resolution in the region of interest of our problem, the outer part of the stellar envelope, than with equal-mass particles.
The resulting minimum particle masses are $5.5\times10^{-7}$\,$M_\odot$, $9\times10^{-7}$\,$M_\odot$, $4.5\times10^{-7}$\,$M_\odot$, $3\times10^{-7}$\,$M_\odot$, and $8.5\times10^{-7}$\,$M_\odot$ for \modelone, \modeltwo, \modelthree, \modelfour, and \modelfive respectively.
For example, our model with $N\simeq3\times10^5$ has an effective resolution in the region of interest similar to a model with $N\simeq4\times10^6$ equal-mass particles.

\begin{table*}
	\centering
	\caption{Relaxed \starsmasher stellar models. $N$ is the number of particles. The values are provided for a specific adopted value of \texttt{nnopt} in each case (see details in \refsec{sec:sph_stellar_models} on how the choice is made). $R_\text{1D}$ is the stellar radius of a 1D \mesa model, $R_\text{SPH}$ is the distance of the outermost SPH particle from the center after the relaxation, $R_{2h}=\max(r_i+2h_i)$ is the distance of the outermost edge of the simulated fluid, $R_\text{edge}$ is the radius within which particles are initially placed in the hcp lattice, and $2h_\text{c}$ and $m_\text{c}$ are the core particles' kernel sizes and masses, respectively. $E_{0.1}$ and $E_{1.0}$ are the binding energies of the envelope above the mass coordinates $1.42$\,$M_\odot$ (top $0.1$\,$M_\odot$) and $0.52$\,$M_\odot$ (top $1$\,$M_\odot$), in $10^{47}$ erg. In the case of our 3D models, the shown energies are from the end of relaxation. All quantities that have dimensions of length are given in $R_\odot$, and those that have dimensions of mass are given in $M_\odot$.}
	\label{tab:rel_models}
    \begin{tabular}{lllllllllll}
        \hline
        Model &
        $N$ &
        \texttt{nnopt} &
        $R_\text{1D}$ &
        $R_\text{SPH}$ &
        $R_{2h}$ &
        $R_\text{edge}$ &
        $2h_\text{c}$ &
        $m_\text{c}$ &
        $E_{0.1}$ &
        $E_{1.0}$ \\
        \hline
\mesa       &          &      & $ 3.05$ &         &         &         &         &         & $ 1.99$ & $ 20.2$  \\
\modelone   & $ 99955$ & $35$ &         & $ 2.93$ & $ 3.30$ & $ 2.70$ & $ 0.24$ & $ 0.16$ & $ 2.01$ & $ 19.9$  \\
\\
\mesa       &          &      & $ 3.72$ &         &         &         &         &         & $ 1.59$ & $ 11.9$  \\
\modeltwo   & $ 99955$ & $27$ &         & $ 3.56$ & $ 4.06$ & $ 3.32$ & $ 0.27$ & $ 0.19$ & $ 1.61$ & $ 11.9$  \\
\modelthree & $200221$ & $57$ &         & $ 3.56$ & $ 3.99$ & $ 3.32$ & $ 0.27$ & $ 0.19$ & $ 1.61$ & $ 11.9$  \\
\modelfour  & $299929$ & $79$ &         & $ 3.57$ & $ 3.99$ & $ 3.32$ & $ 0.26$ & $ 0.19$ & $ 1.61$ & $ 11.8$  \\
\\
\mesa       &          &      & $ 10.0$ &         &         &         &         &         & $ 0.58$ & $ 4.24$  \\
\modelfive  & $ 99955$ & $36$ &         & $ 9.46$ & $ 10.7$ & $ 8.84$ & $ 0.78$ & $ 0.25$ & $ 0.59$ & $ 4.22$  \\
        \hline
    \end{tabular}
\end{table*}

Near the center of the star, using the central density from 1D models in \refeq{eq:m_i} can predict the mass for one particle to be larger than the mass of the star itself.
The reason is that, through the volume allocated for one particle in a 3D model, the density in the 1D model drops significantly.
We therefore assign the mass of the particle located at the center of our 3D models differently.
We call this centrally located particle a `core particle', and we call all other particles `envelope particles'.
A core particle interacts gravitationally but not hydrodynamically with the envelope particles.
We first assign initial masses for all envelope particles and then calculate the core particle mass $m_\text{c}=M_\text{\mesa}-\sum_i^N m_i$, where $M_\text{\mesa}$ is total mass of the 1D \mesa model and $m_i$ is the mass of envelope particle $i$, as calculated with \refeq{eq:m_i}.
We list $m_\text{c}$ for \modelone, \modeltwo, \modelthree, \modelfour, and \modelfive in \reftab{tab:rel_models}.

Following \citet{1992ARA&A..30..543M}, any SPH method calculates any quantity $A(\vect{r})$ at any position $\vect{r}$ as the convolution of $A(\vect{r})$ and some smoothing kernel $W$
\begin{align}\label{eq:sph_calculations}
    A(\vect{r}) &= \int_\text{all space} A(\vect{r}^\prime)W(|\vect{r}-\vect{r}^\prime|,h)d\vect{r}'\ ,
\end{align}
where $h$ is the so-called `smoothing length'. 
Numerically, quantity $A(\vect{r}_i)$ at the location of the $i$'th SPH particle $\vect{r}_i$ is obtained as the following:
\begin{align}\label{eq:sph_calculations_particles}
    A(\vect{r}_i) &= \sum_j^{N_{nb,i}} m_j \frac{A(\vect{r}_j)}{\rho_j} W(|\vect{r}_i-\vect{r}_j|,h_i)\ ,
\end{align}
where $m$ is mass, $\rho$ is mass density, and $N_{nb,i}$ is number of `neighbors'. A neighbor $j$ to particle $i$ is another SPH particle with position $\vect{r}_j$ inside $i$'s kernel.

SPH particles can form close pairs in a so-called `clumping' or `pairing' instability \citep{1981A&A....97..373S} if $W$ has a negative multidimensional Fourier transform \citep{2012MNRAS.425.1068D}.
We therefore adopt the Wendland $C^4$ smoothing kernel, which has a non-negative Fourier transform and thus avoids the clumping instability \citep{2012MNRAS.425.1068D}. 
We scale the kernel to have a compact support of radius $2h$.

To control individual particle number densities, $N_{nb,i}$ is not fixed but is instead dynamically allocated.
Usually the number of neighbors is constrained as $N_{nb,i}=f(\rho_i,C_i)$ where $C_i$ is a constant with dimensions of mass.
The value of $C_i$ is determined at initialization and thus reflects the mass resolution of particle $i$, which is the initial total mass of its neighbors.
Particles from significantly different mass resolutions mix during a stellar merger event and thus $N_{nb,i}=f(\rho_i,C_i)$ becomes inappropriate, as particles may enter a new environment with a larger or smaller average particle mass and thus take on too many or too few neighbors respectively.
We instead adopt the method described in Appendix A of \citet{2010MNRAS.402..105G}, in which a new parameter, \texttt{nnopt}, controls the allocation of $N_{nb,i}$
\begin{align}
    \mathtt{nnopt} &= \sum_j^{N_{nb,i}} G(|\vect{r}_i-\vect{r}_j|,h_i)\ ,
    \label{eq:nnopt}\\
    G(x,h) &\equiv \begin{cases}
        V(4h-4|x-h|,h) & 0\leq x\leq 2h\ , \\
        0 & \text{otherwise}\ ,
    \end{cases}\\
    V(x,h) &\equiv 4\pi \int_0^x x^2 W(x,h)dx\ .
\end{align}
$G(x,h)$ is shown in Figure A1 of \citet{2010MNRAS.402..105G} for the classic cubic-spline smoothing kernel $W$.
\texttt{nnopt} is held constant in time and we set its value at initialization.
For each particle in every time step, $N_{nb,i}$ is iteratively increased or decreased until \refeq{eq:nnopt} is satisfied.

We calculate for both core and envelope particles the gravitational potential estimator \citep{2007MNRAS.374.1347P}
\begin{align}\label{eq:gravitational_potential}
    \hat{\phi}_i = -G\sum_j^{N_{nb,i}} m_j \phi(|\vect{r}_i-\vect{r}_j|,h_j)\ ,
\end{align}
where $G$ is the gravitational constant and $\phi$ is the softening kernel and is related to the smoothing function through
\begin{align}
    W(r_{ij}) &= -\frac{1}{4\pi r_{ij}^2}\frac{\partial}{\partial r_{ij}}\left(r_{ij}^2\frac{\partial \phi}{\partial r_{ij}}\right)\ ,
\end{align}
where $r_{ij}\equiv|\vect{r}_i-\vect{r}_j|$.
A core particle only interacts gravitationally with with other particles, and $\hat{\phi}_i$ is calculated for core particles identically to how $\hat{\phi}_i$ is calculated for envelope particles.
We set the core particle's smoothing length $h_\text{c}$ as the minimum smoothing length of all envelope particles.
We emphasize that $h_\text{c}$ is used only to calculate the softened gravitational potential and is held constant after initialization, unlike envelope particle smoothing lengths which are not held constant and are used to calculate both hydrodynamical and gravitational interactions.
The obtained values of the core particle kernel radii $2h_\text{c}$ and masses $m_\text{c}$ are provided in \reftab{tab:rel_models}.

Our simulated 3D stellar models tend to expand when seeking hydrodynamical equilibrium after initialization, which we adjust for only in models \modelone, \modeltwo, and \modelfive by restricting initial particle positions to be within a sphere of radius
\begin{equation}
    R_{\text{edge}} = R_\text{1D} - \frac{3}{2} \left(1.3 \frac{\mathtt{nnopt}}{N}\right)^{1/3}R_\text{1D}\ ,
    \label{eq:Redge}
\end{equation}
where $R_\text{1D}$ is the radius of the 1D model and $\frac{3}{2} (1.3 \mathtt{nnopt}/N)^{1/3}R_\text{1D}$ is the expected radii of the surface particles' kernels.
For models \modelthree and \modelfour we do not use \refeq{eq:Redge}. 
Instead, we manually restrict the initial particle positions to be within the $R_{\text{edge}}$ value we calculated for \modeltwo.

One of the important quantities that characterizes stellar models, especially during merger interactions, is the binding energy of the envelope. For our spherically-symmetric stars, we calculate the binding energy $E$ above mass coordinate $m(r)$ as \cite[see, e.g.][]{comenv_book_2020}
\begin{equation}
    E(m(r))  = \int_{m(r)}^{M} 
      \left[\frac{Gm(r)}{r} - u(r)\right] dm\ ,
    \label{eq:Ebind}
\end{equation}
where $u(r)$ is the specific internal energy.
We obtain values of $E(m(r))$ both for our 1D and 3D models and name each as $E_\mesa(m(r))$ and $E_\text{SPH}(m(r))$, respectively.
To calculate $E_\text{SPH}(m(r))$ we first sort the particles by distance to the core particle $r_i$ and calculate the particle mass coordinate
\begin{align}
    m(r_i) &= \sum_{j=c}^i m_j,
\end{align}
and the envelope binding energy
\begin{align}\label{eq:EbindSPH}
    E_{\text{SPH}}(m(r)) = \sum_{m(r_i) \geq m(r)} \left[\frac{Gm(r_i)}{r_i} - u_i\right]m_i\ .
\end{align}
We do not use $\hat{\phi}_i$ to calculate the gravitational potential energy in \refeq{eq:EbindSPH}.

\subsection{Stellar model relaxation}\label{sec:relaxation}%
We let our individual \starsmasher models settle to hydrostatic equilibrium over time, or `relax'.
During relaxation, we use artificial viscosity for  hydrodynamical accelerations. 
However, to avoid artificial heating, we do not use artificial viscosity when calculating the change in specific internal energy $\dot{u}_i$.
The change in specific internal energy is calculated by the method detailed in \refeqlabel~\refeqstyle{A18} in \refapplabel~A of \citet{2010MNRAS.402..105G}.

We obtain envelope energies:
\begin{equation}
    U = \sum_i^N m_i u_i\ , \  W_g = \frac{1}{2} \sum_i^N m_i \hat{\phi}_i\ , \  T = \frac{1}{2} \sum_i^N m_i v_i^2\ ,
    \label{eq:grav_pot_SPH}
\end{equation}
where $U$ is the internal energy of the envelope, $W_g$ is the gravitational potential energy of the envelope, $T$ is the kinetic energy of the envelope, and $v_i$ is particle velocity. 
We multiply $W_g$ by a factor of $1/2$, as shown in \refeq{eq:grav_pot_SPH}, to correct for the double counting of particles in \refeq{eq:gravitational_potential}. We calculate $U$ by summing over all envelope particles (excluding core particles), whereas we calculate $W_g$ and $T$ by summing over all particles, core particles included.

We trace the change in $U$ and $W_g$ with time by fitting each as a simple harmonic oscillator function with diminishing amplitude and a mean which varies as a power law
\begin{equation}
    f(t) = p_1 t^{p_2} \sin(2\pi p_3t + p_4) + p_5 t^{p_6}\ ,
    \label{eq:relaxation_fitting}
\end{equation}
where $p_n$ are fitting parameters.
We solve for all $p_n$ in \refeq{eq:relaxation_fitting} for both $U$ and $W_g$ separately in the final half (by time) of each of our relaxations using a Levenberg--Marquardt least-squares fit \citep{1978LNM...630..105M} as implemented in \texttt{scipy} version 1.2.1.
The resulting amplitudes in both $U$ and $W_g$ at the final time are $\lesssim 10^{45}$\,ergs. 
Each of our fits have a coefficient of determination (commonly called the `R-squared' value) between $0.986$ and $0.998$.
The values of $U$ and $W_g$ in all our models are about $10^{47}$\,ergs in magnitude in the region of our interest, top $0.1$\,$M_\odot$, and an order of magnitude larger for the entire envelope.
Hence, all our models are relaxed until $U$ and $W_g$ vary on the order of or less than 1\,per~cent.

\subsection{Choosing the best relaxed models}\label{sec:matching_the_mesa_model}%
In the case of a merger, only a small fraction of the envelope participates in the dynamical interaction during the dynamical part of the event and forms the outflow.
Specifically, in the case of V1309~Sco, only the outer $0.04$--$0.08$\,$M_\odot$ is expected to be in the outflow \citep{2014ApJ...786...39N}.
Matching the energy profiles in the outermost layers is hence more important in the study of a merger than in the case of a complete common envelope ejection event, where the fate of the entire envelope must be decided.
Even more challenging is to reproduce the energy in the visible surface layer, which is, at first, a boundary layer, and also contains only a tiny fraction of the overall energy, which is comparable with the overall convergence error.

For our base models, we analyze how well $E_{\text{SPH}}(m(r))$ matches $E_{\text{\mesa}}(m(r))$ in the outermost $0.1$\,$M_\odot$ of our relaxed stellar model envelopes to the outermost $0.1$\,$M_\odot$ of the initial 1D models.
To quantify, we calculate the normalized relative percentage error between the binding energies of our 3D models and their corresponding 1D models
\begin{equation}
    \delta E(m(r)) = \frac{E_{\text{\mesa}}(m(r))-E_\text{SPH}(m(r))}{E_{\text{\mesa}}(1.42\,M_\odot)} \times 100\ ,
    \label{eq:delta_E}
\end{equation}
where $E_{\text{\mesa}}(1.42\,M_\odot)$ is the binding energy at the shell with mass coordinate closest to but not exceeding $1.42$\,$M_\odot$.

\begin{figure}
    \centering
    \includegraphics{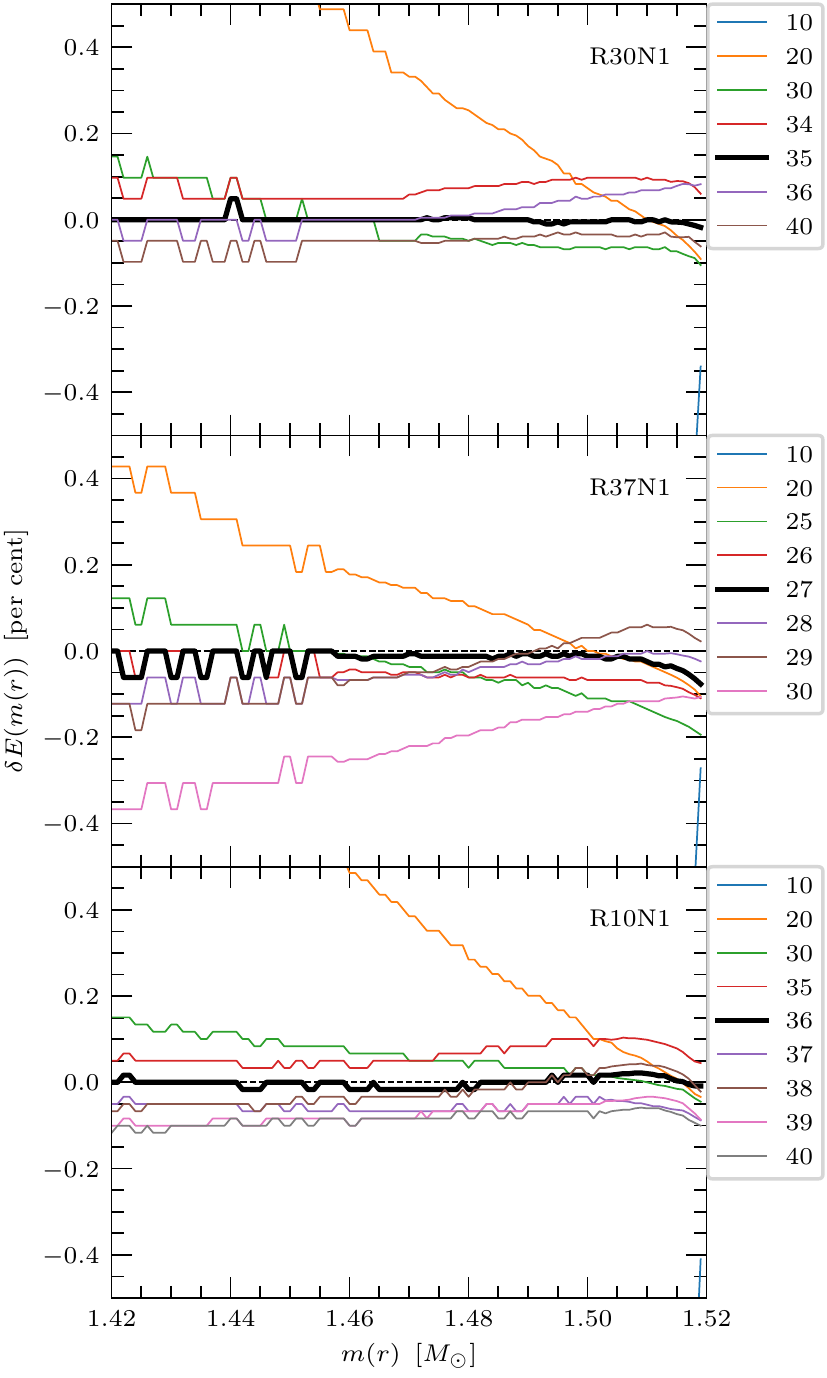}
    \caption{We search for the optimal \texttt{nnopt} values (black lines) for \modelone (top), \modelthree (middle), and \modelfive (bottom) relaxed \starsmasher models using the relative percentage errors in integrated total energies $\delta E(m(r))$ between our \starsmasher models and their initial 1D \mesa models, normalized to mass coordinate $m(r)=1.42$\,$M_\odot$. 
    We find the values $\mathtt{nnopt}=35$, $\mathtt{nnopt}=27$, and $\mathtt{nnopt}=36$ best match each respective profile. 
    We provide in \reftab{tab:table_nnopts} the sum of squared residuals for each \texttt{nnopt} value shown.}
    \label{fig:optimal_nnopt}
\end{figure}

\begin{table}
	\centering
	\caption{The sum of squared residuals (see \refeqp{eq:residuals}) for relaxed 3D models obtained with different  \texttt{nnopt} values (see also \reffig{fig:optimal_nnopt}). The \texttt{nnopt} values that give the minimum $\sum\left(\text{resid}\right)^2$ are 35, 27, and 36 for \modelone, \modeltwo, and \modelfive respectively.}
	\label{tab:table_nnopts}
    \begin{tabular}{llllll}
        \hline
        \multicolumn{2}{c}{\modelone} &
        \multicolumn{2}{c}{\modeltwo} &
        \multicolumn{2}{c}{\modelfive} \\
        \texttt{nnopt} & $\sum \left(\text{resid}\right)^2$ &
        \texttt{nnopt} & $\sum \left(\text{resid}\right)^2$ &
        \texttt{nnopt} & $\sum \left(\text{resid}\right)^2$ \\
        \hline
        10 & $2.69\times10^4$ & 10 & $3.03\times10^4$ & 10 & $3.62\times10^4$ \\
        20 & 20.8 & 20 & 4.56 & 20 & 25.9 \\
        30 & 0.44 & 25 & 0.73 & 30 & 0.60 \\
        34 & 0.55 & 26 & 0.31 & 35 & 0.45 \\
        35 & 0.01 & 27 & 0.08 & 36 & 0.01 \\
        36 & 0.15 & 28 & 0.47 & 37 & 0.33 \\
        40 & 0.33 & 29 & 0.63 & 38 & 0.15 \\
           &      & 30 & 5.72 & 39 & 0.58 \\
           &      &    &      & 40 & 0.74 \\
        \hline
    \end{tabular}
\end{table}

We search for the best fit by varying \texttt{nnopt} for \modelone, \modeltwo, and \modelfive until we find the minimum sum of squared residuals over 100 equally-sized mass bins in the outermost $0.1$\,$M_\odot$:
\begin{align}\label{eq:residuals}
    \sum(\text{resid})^2 \equiv \sum_{m(r)\geq1.42\,M_\odot} \left[\delta E(m(r))\right]^2\ .
\end{align}
For each model, we start with $\mathtt{nnopt}=10$ and increase \texttt{nnopt} until we can get the best match for the energy profile. 
In \reffig{fig:optimal_nnopt} we demonstrate the qualitative change of the profiles with \texttt{nnopt}.
In \reftab{tab:table_nnopts} we list the quantitative estimates. 
We choose the \texttt{nnopt} values for which $\sum(\text{resid})^2$ is minimized in the top $0.1$\,$M_\odot$ of the envelope.

In \reftab{tab:rel_models} we show the binding energies of the top $0.1$\,$M_\odot$ of the envelope $E_{0.1}$ and of the top $1$\,$M_\odot$ as $E_{1.0}$ for our \mesa models and for our fully relaxed 3D models with best \texttt{nnopt}.
We find that values of $E_{0.1}$ and $E_{1.0}$ in our 3D models are within $0.9$--$1.8$\,per~cent of their associated \mesa models.

\subsection{Chosen relaxed models}\label{sec:relaxed_models}%
In \reftab{tab:rel_models}, we show the radii and energies of our best relaxed 3D models.
We list there two ways to determine a size of a star in SPH code.
$R_\text{SPH}$ is the radius within which all SPH particles are located, while $R_{2h}$ is the distance from the origin of the outermost particles' influences.
\begin{align}
   R_\text{SPH} &\equiv \max\left(r_i\right)\ , \\
   R_{2h} &\equiv \max\left(r_i + 2h_i\right)\ .
\end{align}
\noindent There is no hydrodynamical influence by any SPH particle outside of $R_{2h}$; matter density is zero there.
The radii of our 1D models, $R_\text{1D}$, is intrinsically the photospheric radius, where the optical depth is $\tau=2/3$.
The 1D model is not obtained above the photosphere, to where the star has a zero density.
The 1D and 3D models' radii therefore should not be directly compared.

Each of our surface SPH particles has a very large own optical depth $\tau_i\gg 2/3$ (we will discuss how we obtain $\tau_i$ in \refsec{sec:calculating_flux_in_SPH}).
We can measure the simplistic `optical depth profile' of an isolated SPH particle by integrating inwards toward its center, from its surface located at $2h_i$ to some distance from the particle center, and then scale it with respect to the total optical depth of the SPH particle.
The optical depth is a strong function of the distance to the center of the particle.
With the kernel we use and with opacity proportional to density, $\tau_i(z)\approx 2\times 10^{-3} \tau_i$ at distance $z=h_i$ from the center, and $\tau_i(z)\approx 10^{-11} \tau_i$ at distance $z=1.8 h_i$. 
Therefore, the `photospheric' radius of 3D models is located somewhere between $R_\text{SPH}<R_\text{phot}<R_{2h}$, and likely further limited to the region $R_{h}<R_\text{phot}<R_{2h}$, where $R_{h}=(R_\text{SPH}+R_{2h})/2$.
This makes it close to $R_\text{1D}$.

%% file: Section3.tex
\section{Calculating Flux in SPH}\label{sec:calculating_flux_in_SPH}%
We choose to frame the problem of calculating the radiative flux from SPH simulations in the context of so-called `ray tracing', as has been done in previous methods \citep[see, e.g.][among others]{2008MNRAS.386.1931A,2013MNRAS.434..748A,2015MNRAS.449..243N,2017ApJ...850...59P,2017ApJS..229...36G}.
We define a `ray' as a straight line which intersects with the position of an observer who is infinitely far away.
Along any given ray we assume that the physical fluid, which is represented by the SPH particles, can instead be represented by a thick slab with equivalent physical properties to the fluid, such as temperature and density, at all levels.
We let the slab surface be normal to the ray and define the radiative flux
\begin{equation}
    F^+ = F_0^+(\tau) e^{-\tau} + \pi \int_0^\tau S(\tau^\prime) e^{-\tau^\prime}  d\tau^\prime\ .
    \label{eq:flux1}
\end{equation}
Here $S(\tau^\prime)$ is the source function and $F_0^+(\tau)$ is the radiative flux with which the slab is illuminated at optical depth $\tau$ that is measured from the observer:
\begin{equation}
    \tau = \int_0^z \rho(z^\prime) \bar{\kappa}(z^\prime) dz^\prime\ .
    \label{eq:optical_depth}
\end{equation}
Here $\rho$ is the mass density, $\bar{\kappa}$ is the mean opacity, and $z$ is the distance measured along the ray in the direction from an arbitrary observer, and into the object. 
We define $z$ as zero where $\tau=0$, which is where the mass density of the stellar material along the ray becomes non-zero.
We calculate $\bar{\kappa}$ using the Rosseland mean opacity table from \mesa \citep[see \refseclabel4.3 of][]{Paxton2011} and a Planck mean opacity table from \citet{2003A&A...410..611S}, blended together with cubic spline interpolation between 990\,K and 1100\,K.

At large $\tau$ in stellar interiors, local thermodynamic equilibrium is usually closely satisfied.
The flux we integrate does not necessarily form at a large $\tau$, but we still adopt that the temperature of the local matter and the radiation field it creates can be closely described by the Planck blackbody function.
In that case, the source function is
\begin{equation}
    S(\tau^\prime) = \frac{\sigma_\text{SB}}{\pi} T(\tau^\prime)^4 \ ,
    \label{eq:blackbody_assumption}
\end{equation}
where $\sigma_\text{SB}$ is the Stefan--Boltzmann constant, and $T(\tau^\prime)$ is temperature at the optical depth $\tau^\prime$.

The effective temperature $T_\text{eff}$ is usually defined in stellar physics as a temperature that would create the observed flux if the body were a blackbody, and we follow the same convention
\begin{equation}
    F^+ = \sigma_{\text{SB}} T_\text{eff}^4\ .
    \label{eq:teff_flux}
\end{equation}
We can find $T_\text{eff}$ along any given ray by applying {\refeqlabel}s~\refeqstyle{\ref{eq:blackbody_assumption}} and \refeqstyle{\ref{eq:teff_flux}} to \refeq{eq:flux1}:
\begin{equation}
    T_\text{eff} = \left[\frac{1}{\sigma_\text{SB}} F_0^+(\tau) e^{-\tau} + \int_0^\tau T(\tau^\prime)^4 e^{-\tau^\prime} d\tau^\prime \right]^{1/4}
    \label{eq:Teff}
\end{equation}
along each ray, in the direction from the observer.

As can be seen from \refeq{eq:Teff}, the contribution of local matter into the formation of the outgoing radiative flux drops down with optical depth as $e^{-\tau}$.
In what follows, we adopt that an optically thick particle has an optical depth $\tau_\text{th}=20$ or more.
To determine the total individual optical depths of each SPH particle that is intersected by a ray, we use the value of a particle's optical depth in isolation:
\begin{equation}
    \tau_i \approx m_i \bar{\kappa}_i \int_0^{2h_i} W(z,h_i) dz = \frac{m_i \bar{\kappa}_i}{h_i^2} C_\tau \ .
    \label{eq:particle_tau}
\end{equation}
Here $\bar{\kappa}_i$ is the particle mean opacity (obtained from our blended opacity tables), and $C_\tau$ is a dimensionless constant and has a value of $C_\tau = 55/(48\pi) \approx 0.36$ for the Wendland $C^4$ function.
We emphasize that the optical depth as calculated with \refeq{eq:particle_tau} is distinct from the optical depth as calculated by ray tracing, as ray tracing is not limited to a single particle in isolation.
We use \refeq{eq:particle_tau} only to inform our choice of methods in calculating $T_\text{eff}$ for a ray (see \refsec{sec:calculating_effective_temperatures_on_a_grid}).

We discuss in \refsec{sec:ray_tracing} our ray tracing method and show that a standard ray tracing technique is inappropriate for calculating $T_\text{eff}$ for rays which enter an optically thick particle early during integration.
In \refsec{sec:envelope_fitting} we introduce our `envelope fitting' method which we created to obtain $T_\text{eff}$ for rays that cross optically thick SPH particles.
We describe in \refsec{sec:calculating_effective_temperatures_on_a_grid} how we calculate $T_\text{eff}$ of our stars as a whole, and compare the outcomes of the ray tracing method and of the envelope fitting method.

\subsection{Ray tracing}\label{sec:ray_tracing}%
\begin{figure}
	\centering
	\includegraphics{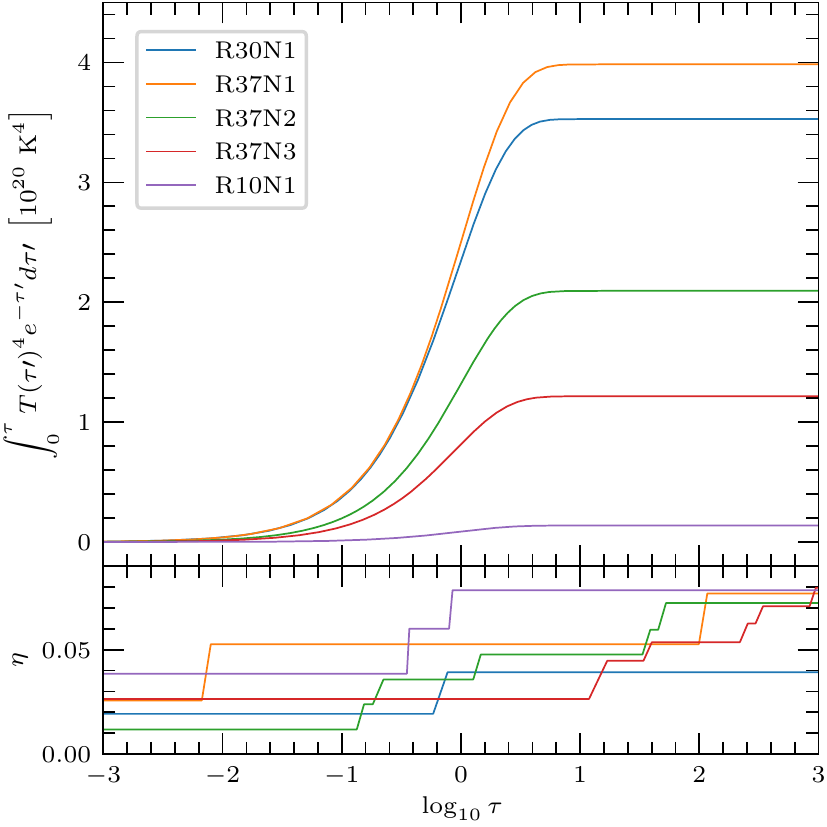}
    \caption{We show in the top panel the integral in \refeq{eq:Teff} (solid lines) at optical depth $\tau$ along a single ray traced through each of our relaxed 3D models \modelone (blue), \modeltwo (orange), \modelthree (green), \modelfour (red), and \modelfive (purple). In the bottom panel we show for the same rays the SPH ray tracing relative accuracy metric $\eta$ from \refeq{eq:eta}.}
    \label{fig:tau_with_Teff_and_eta}
\end{figure}

We define $\vect{r}_z$ as the 3D position of the point on the ray, which is determined uniquely by the distance $z$.
We remind the reader that $\vect{r}_z$ is a 3D quantity, whereas $z$ is a 1D quantity.
If $\vect{r}_z$ is covered by SPH particle kernels, we can use a variation of \refeq{eq:sph_calculations_particles} to find the local temperature and density:
\begin{align}
    T(\vect{r}_z) &= \frac{1}{\rho(\vect{r}_z)} \sum_j^{N_{nb}(\vect{r}_z)} m_j T_j W(|\vect{r}_z-\vect{r}_j|,h_j)\ ,\label{eq:Tz} \\
    \rho(\vect{r}_z) &= \sum_j^{N_{nb}(\vect{r}_z)} m_j W(|\vect{r}_z-\vect{r}_j|,h_j)\ .\label{eq:rhoz}
\end{align}
Here $N_{nb}(\vect{r}_z)$ is the number of kernels that overlap at $\vect{r}_z$.
Then, for any given ray, we can find the effective temperature $T_\text{eff,ray}$, using \refeq{eq:Teff}, where for the optical depth $\tau$ we use \refeq{eq:optical_depth}.

Typically the value of $N_{nb,i}$ is controlled by the SPH code by adjusting the particle smoothing lengths so that the accuracy of the approximation in \refeq{eq:sph_calculations_particles} is constrained.
{\Refeqlabel}s~\refeqstyle{\ref{eq:Tz}} and \refeqstyle{\ref{eq:rhoz}} differ from \refeq{eq:sph_calculations_particles} in that $N_{nb}(\vect{r}_z)$ is not controlled in any way.
As such, $N_{nb}(\vect{r}_z)$ can be as low as one near the SPH simulation boundary such that the distribution of temperature and density from {\refeqlabel}s~\refeqstyle{\ref{eq:Tz}} and \refeqstyle{\ref{eq:rhoz}} is entirely described by the smoothing kernel $W$.

We can quantify the accuracy of our ray tracing method in relation to the accuracy of the SPH simulation, with a new parameter which we hereafter call the `SPH ray tracing relative accuracy metric':
\begin{align}\label{eq:eta}
    \eta &\equiv \frac{N_{nb}(\vect{r}_z)}{\langle N_{nb,i}\rangle_{\vect{r}_z}}\ ,
\end{align}
where $\langle N_{nb,i}\rangle_{\vect{r}_z}$ is the average number of neighbors over all particles which constitute $N_{nb}(\vect{r}_z)$
\begin{align}
    \langle N_{nb,i}\rangle_{\vect{r}_z} &\equiv \frac{1}{N_{nb}(\vect{r}_z)}\sum_i^{N_{nb}(\vect{r}_z)} N_{nb,i} \ .
\end{align}
Here the sum is over all particles whose kernels overlap $\vect{r}_z$.

The value of $N_{nb,i}$ is an indicator for the accuracy of the quantities calculated by the SPH simulation, such as $T_i$ and $\rho_i$ and, likewise, the value of $N_{nb}(r_z)$ is an indicator for the accuracy of the quantities calculated by ray tracing, namely $T(\vect{r}_z)$ and $\rho(\vect{r}_z)$.
If $\eta\sim 1$, then the number of terms $N_{nb}(\vect{r}_z)$ used in the summations of {\refeqlabel}s~\refeqstyle{\ref{eq:Tz}} and \refeqstyle{\ref{eq:rhoz}} is comparable to the number of terms $N_{nb,i}$ used in \refeq{eq:sph_calculations_particles} for a nearby particle $i$, and the accuracy of $T(\vect{r}_z)$ and $\rho(\vect{r}_z)$ is comparable to that of $T_i$ and $\rho_i$ as calculated by the SPH code.
For $\eta\ll1$, which occurs near the very surface of the mass distribution, the accuracy of {\refeqlabel}s~\refeqstyle{\ref{eq:Tz}} and \refeqstyle{\ref{eq:rhoz}} is poor because very few particles contribute to the interpolation and those particles are asymmetrically distributed around $\vect{r}_z$.

We trace a ray through each of our relaxed 3D models with uniform integration step sizes $\Delta z = 0.002$\,$R_\odot$ and find that contributions to $T_\text{eff,ray}$ become negligibly small at $\tau \gtrsim 10$, as shown in the top panel of \reffig{fig:tau_with_Teff_and_eta}.
When we start integrating $\tau$ along the ray starting from $z=0$, while the optical depth is still small ($\tau$ about a few or less), we find that $\eta<0.1$. 
The region of $\tau\sim1$ contributes the most to the finally obtained value of $T_\text{eff,ray}$ and, although the best accuracy is achieved by integrating to $\tau \gtrsim 10$, such regions do not contribute significantly to the final value of $T_\text{eff,ray}$.

\subsection{Envelope fitting method}\label{sec:envelope_fitting}%
\begin{figure}
    \centering
    \includegraphics{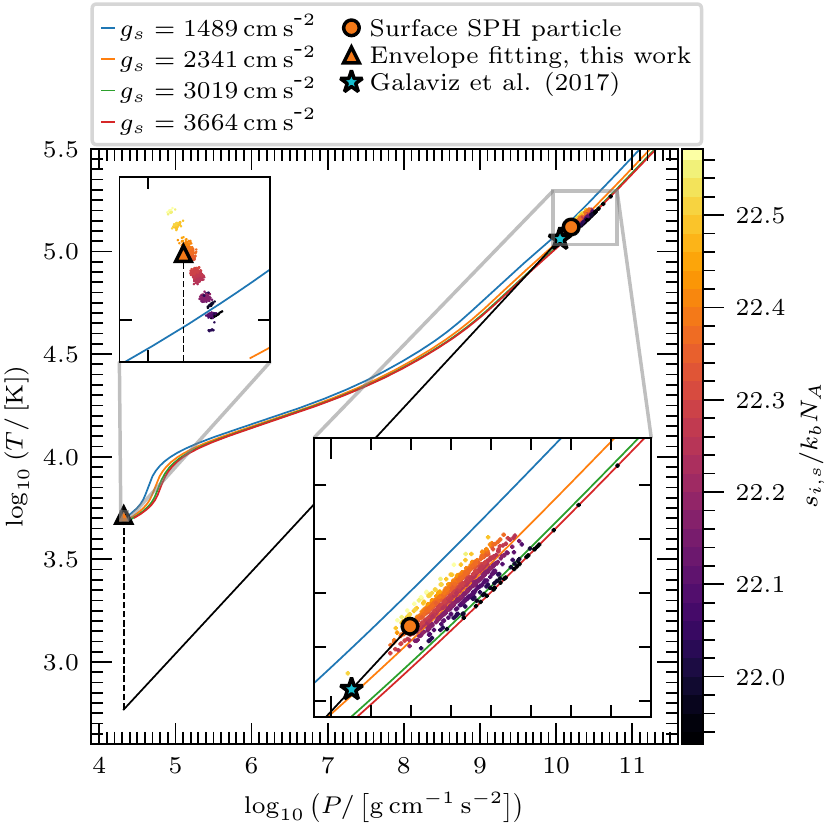}
    \caption{The surface particles (dots) in our relaxed 3D model \modeltwo form $P,\,T$ layers corresponding to their specific entropy values $s_{i,s}$, which we obtain from our TEOS, and scale by $k_bN_A$, where $k_b$ is Boltzmann's constant and $N_A$ is Avogadro's number. We show our 1D \mesa model for \modeltwo (green line), and the same model at various stages of stellar evolution when the surface gravitational acceleration $g_s$ is $\sim20$\,per~cent larger ($R=3.37$\,$R_\odot$; red line), $\sim20$\,per~cent smaller ($R=4.22$\,$R_\odot$; orange line), and $\sim50$\,per~cent smaller ($R=5.29$\,$R_\odot$; blue line). We calculate the surface particles' $T_\text{eff,env}$ and $P_\text{ph}$ (triangles) using our envelope fitting method from \refsec{sec:envelope_fitting}. For a single chosen surface particle (enlarged and outlined symbols), we show the $T_\text{eff}$, $P_\text{ph}$ at the outer edge of the kernel as calculated by a stratified temperature distribution (cyan star) and trace that distribution (solid black line) to $P_\text{ph}$ (dashed black line).}
    \label{fig:PT_with_Galaviz}
\end{figure}

We present here our method for obtaining flux from optically thick particles in SPH simulations by fitting pre-calculated solutions for stellar envelopes.

For optically thick envelopes, stellar envelope profiles form families of unique solutions in the pressure--temperature plane $P$--$T$,  where the solutions are determined by entropy of the interiors and the pair of values formed by surface gravity $g_s$ and $T_\text{eff}$ \citep[\refseclabel11.3 in ][see also envelope profiles shown in \reffig{fig:PT_with_Galaviz}]{2012sse..book.....K}
\footnote{In \cite{2012sse..book.....K}, the solutions of folded stellar structure equations for the envelope were uniquely defined by the two quantities, $B$ and $C$.
The quantity $B\propto L/M$, which we translate into $B\propto T_\text{eff}^4/g_s$. 
The quantity $C$, named as "a constant of integration" in \cite{2012sse..book.....K}, together with $B$ provides a relation between $P$ and $T$.
This relation gives entropy dependence to the solution families.
We do not rewrite here the equations in analogy of \cite{2012sse..book.....K}, but we assume that the parameters $B$ and $C$ which the envelope solution families depend on  can be replaced by $g_s$, $T_\text{eff}$ and the entropy from the actual stellar envelope solutions.}. 
The uniqueness of families of solutions is a consequence of the folding of stellar structure equations into one parametric equation.
The limitation however is that this uniqueness is only valid if the envelope is in a state of both hydrostatic and thermal equilibrium (the latter implies that $L$ does not change throughout the considered part of the envelope).
The end point of the envelope solution is where the combination of $P$ and $T$ also satisfies an adopted relation between the pressure at the photosphere $P_\text{ph}$ and the effective temperature of a stellar envelope $T_\text{eff,env}$.
The location of an envelope profile in $P$--$T$ plane can also be described by the envelope's entropy.
If the entropy is known, and there is a relation between effective temperature and pressure at the surface, then only one unique envelope profile is possible for each value of surface gravity.
In addition, those solutions are indirect functions of the adopted chemical composition (implied to be uniform along the $P$--$T$ envelope profile) and used opacity tables.

We extracted the actual stellar envelope profiles, as calculated by \mesa, for stars of the same metallicity as our star, V1309~Sco.
We first built 1D zero-age main sequence stellar models of 20 various initial masses, from $0.8$ to $30$\,$M_\odot$, using \mesa, and then evolved them throughout the HR diagram, stopping before the dredge-up  could significantly change the surface composition; no star was evolved beyond the end of He core burning. The envelope solutions were extracted for a wide range of surface gravities, $-4 \leq \log_{10}(g_s/[\text{cm}\,\text{s}^{-2}]) \leq 4.6$, in increments of $0.01$\footnote{A comparison of results was made, and choosing between the increments of 0.01 and 0.1 had no effect on the results presented in this paper in \S4, but this independence on the increment can not be guaranteed for all plausible uses.}. 
Specifically during the the post-processing of the simulations present in this paper, we used a subset of the larger table for the range $2 \leq \log_{10}(g_s/[\text{cm}\,\text{s}^{-2}]) \leq 4$. During the processing, for each particle with some specific value of $g_s$, the envelope profile that has the entropy closest to that of that SPH particle is chosen.

We do not interpolate from the stellar envelope profiles directly, but instead assume that it is the shape of the solution that is preserved, and then interpolate between the shapes.
We parameterize the envelope solutions by recording the `effective' slope at each point of the envelope
\begin{equation}
    \nabla (P,T,g_s) \equiv  \frac{\log_{10}(T_\text{eff,env}/T)}{\log_{10}(P_\text{ph}/P)}\ ,
    \label{eq:nabla}
\end{equation}
as a function of temperature between $3.3 \leq \log_{10}(T/[\text{K}]) \leq 6.0$, in increments of $0.01$, for each $g_s$.
Note that $\nabla (P,T,g_s)$ is not a local derivative, but just the quantity that allows for a quick recovery of the actual stellar envelope solutions for SPH particles, as described below.
When we process our 3D models, we use the table of $\nabla (P,T,g_s)$, not the extracted stellar envelope profiles.

For an SPH particle $i$, we adopt the Eddington approximation
\begin{equation}
    P_\text{ph} \approx \frac{2}{3} \frac{g_i}{\bar{\kappa}_s}\ ,
    \label{eq:env_fit_Pph}
\end{equation}
where $g_i$ is the local gravitational acceleration of particle $i$ and $\bar{\kappa}_s$ is the mean opacity at the surface (it is itself a function of $T_\text{eff}$ and $P_\text{ph}$). 
We compute $\bar{\kappa}_s$ using the low-temperature opacity table from \citet{Ferguson2005}, which is included in the version of \mesa we use (9793) with the file name, `\texttt{lowT\_fa05\_gs98\_z0.02\_x0.7.data}'.
For consistency, all the stellar models that we used to extract the envelope profiles were evolved using the same Eddington approximation.

Further, for each SPH particle $i$ that has central values of $P_i$ and $T_i$, we obtain $\nabla$ from our created $\nabla$ table, interpolating for $g_s$ and temperature while suing the envelope profile that has the closest entropy to the considered SPH particle.
We obtain the `fitted' pressure at the surface
\begin{equation}
    P = P_i\left(\frac{T}{T_i}\right)^{1/\nabla}\ ,
    \label{eq:env_fit_P}
\end{equation}
where $P_i$ and $T_i$ are the pressure and temperature of particle $i$.
Then we solve for the `fitted' effective temperature $T_\text{eff,env}$ with the condition that $P=P_\text{ph}$: in this way, both {\refeqlabel}s~(\ref{eq:env_fit_Pph}) and (\ref{eq:env_fit_P}) are simultaneously satisfied.

We show the values of $P_\text{ph}$ and $T_\text{eff,env}$ for the surface particles from our relaxed 3D model \modeltwo in \reffig{fig:PT_with_Galaviz}.
The $P_\text{ph}$ and $T_\text{eff,env}$ are slightly different from that of our 1D base model due to small variations in entropy ($T_i$ and $P_i$ combinations)  and $g_i$ among the surface particles.
For a single chosen surface particle, we compare our result to that of the stratified temperature distribution from \refseclabel3.5 of \citet{2017ApJS..229...36G}, as shown by the cyan star in \reffig{fig:PT_with_Galaviz}.
In that approach, the effective temperature is calculated by assuming an ideal gas with adiabatic index $\gamma=5/3$ such that $T=T_i(r_\text{ph}/r)^{4/3}$, where $r$ is the particle location and $r_\text{ph}$ is the location of the surface of its kernel.
The result is a $T_\text{eff}$ roughly an order of magnitude higher than our $T_\text{eff,env}$, and at a photospheric pressure about 6 orders of magnitude higher.

\subsection{Calculating effective temperatures on a grid}\label{sec:calculating_effective_temperatures_on_a_grid}%
\begin{figure}
	\centering
	{%
    \setlength{\fboxsep}{0pt}%
    \setlength{\fboxrule}{0.5pt}%
	\fbox{\includegraphics[width=\dimexpr\columnwidth-\fboxrule]{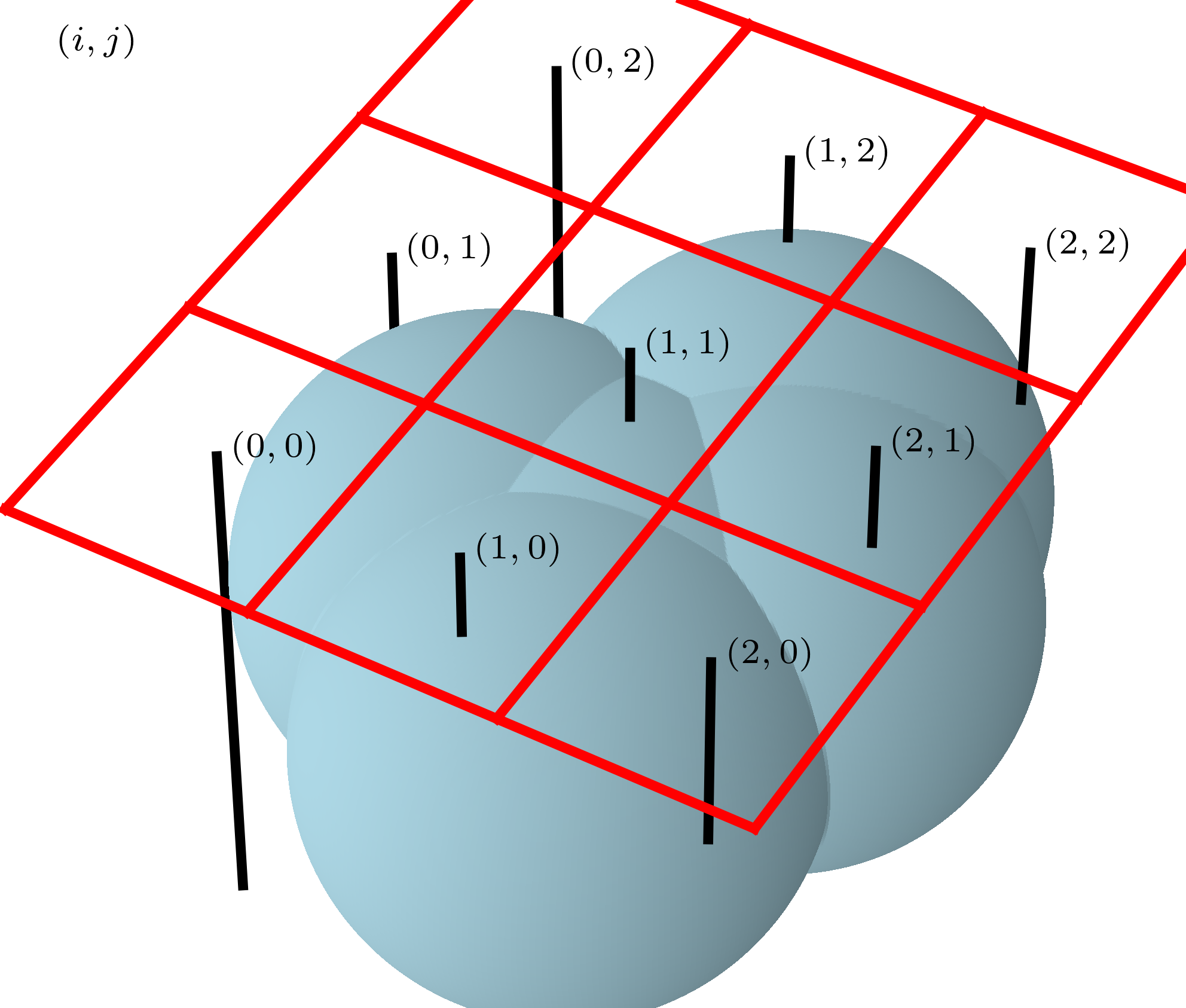}}
	}%
    \caption{A sketch of a $3\times3$ grid (red lines) with rays (black lines; indices indicated by $(i,j)$ in black text) traced into a grouping of SPH particles (blue spheres). To obtain the flux, we can calculate $T_\text{eff}(i,j)$ for each ray and fill the physical area associated with the grid cell $A(i,j)$. For each ray, the integration distance $z$ equals 0 at the location where the ray first enters an SPH particle, closest to the grid.}
    \label{fig:grids}
\end{figure}

To create an image, we start by choosing a direction at which an observer is located.
Then we take a plane which is normal to the direction of the observer, and is just outside of the domain that contains all SPH fluid between the object and the observer.
We project our 3D object on this chosen plane and subdivide the area of the plane into cells to form a grid.
Each grid cell is then considered to be associated with a single ray that extends into the simulated fluid in the direction normal to the chosen plane, with $z=0$ set to the surface of the particle kernel closest to the observer along that ray.
We show an example grid for a chosen collection of SPH particles in \reffig{fig:grids}.

\begin{figure}
	\centering
    {%
    \setlength{\fboxsep}{0pt}%
    \setlength{\fboxrule}{0.5pt}%
	\fbox{\includegraphics[width=\dimexpr\columnwidth-\fboxrule]{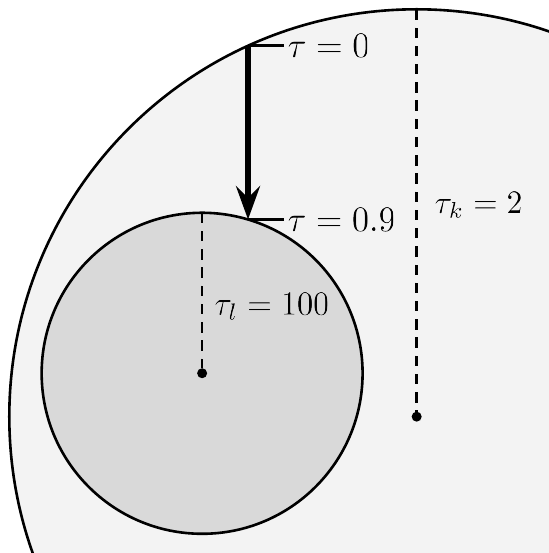}}
	}%
    \caption{A 2D sketch of two SPH particle kernels in which particle $k$ (light grey) has a neighbor, particle $l$ (dark grey), each with example intrinsic optical depths (see \refeqp{eq:particle_tau}) $\tau_k=2$ and $\tau_l=100$. We show an example ray (black arrow) traced through the particle kernels. Here, we set the condition for optical thickness to $\tau_\text{th} = 20$ such that we perform envelope fitting only for particle $2$.}
    \label{fig:Tau_LOS}
\end{figure}

In an arbitrary SPH simulation of a stellar merger event, particles along each ray can have significantly different $\tau$.
As before, we adopt that an optically thick particle has an optical depth $\tau_\text{th}=20$ or more.
Let us consider an example with two particles, `$k$' and `$l$', with $\tau_k=2$ and $\tau_l=100$ correspondingly (see \reffig{fig:Tau_LOS}).
If we begin tracing a ray at the surface of the kernel of particle $k$, where $\tau=0$, we may enter the kernel of particle $l$ after the integrated $\tau$ is only a fraction of $\tau_k$.
To take this into account, for any given ray, we calculate $T_\text{eff}$ in accordance to the following situations:
\begin{enumerate}[(a)]
    \item \label{envfit_off} All particles that intersect the ray $(i,j)$ are optically thin, $\tau_k<\tau_\text{th}$.
    We use only ray tracing, with integration limit $\tau=20$.
    
    \item \label{envfit_on} The first particle intersecting along the ray (surface particle) has $\tau_k>\tau_\text{th}$.
    We use only envelope fitting.

    \item \label{both} In tracing the ray $(i,j)$, contributions to $\tau$ are first made by $\tau_k < \tau_\text{th}$ particles.
    Then, at some $\tau$, the ray enters the kernel of a particle $l$ which has $\tau_l \geq \tau_\text{th}$.
    We use envelope fitting for the particle $l$ to obtain $T_{\text{eff,env},l}$, and find the effective temperature for the ray using \refeq{eq:Teff} in the following form: 
    \begin{equation}
        T_\text{eff} (i,j) = \left[T_{\text{eff,env},l}^4 e^{-\tau} + \int_0^\tau T(\tau^\prime)^4 e^{-\tau^\prime} d\tau^\prime \right]^{1/4}\ .
    \end{equation}

    \item \label{no_fluid} There are no intersecting particles on the ray. 
    No contribution is made to the total outgoing flux. We set $T_\text{eff}=0$ for this ray.
\end{enumerate}
We emphasize that in this work we use only \ref{envfit_on} and \ref{no_fluid} for all rays in all our 3D models, as all surface particles have $\tau_i\gg\tau_\text{th}$ (see \refsec{sec:results} for details).
We plan to use \ref{envfit_off} and \ref{both} in future work.

After $T_\text{eff}$ is obtained for all rays on the 2D grid, 
we find the luminosity in the direction of the observer 
\begin{equation}
    L = \sum_{i,j} 4 \sigma_\text{SB} A(i,j) T_\text{eff}(i,j)^4\ ,
    \label{eq:grid_flux}
\end{equation}
where $A({i,j})$ is the physical area of a single cell, which is the same for all cells on the grid. The leading factor of $4$ arises from the spherical symmetry assumption made by the observer (in reality, true total luminosity can be different).
We refine our grid resolution until the obtained value of luminosity does not change with the next level of refinement by more than 1\,per~cent.

%% file: Section4.tex
\section{Results of Flux Calculations}\label{sec:results}%
\begin{table*}
	\centering
	\caption{We show for our relaxed 3D models in \reftab{tab:rel_models} the effective temperature and luminosity of our base 1D models, $T_{\text{eff},1D}$ and $L_\text{1D}$, and the average effective temperature and luminosity from all rays for which (\ref{no_fluid}) does not apply, $\langle T_\text{eff}\rangle$ and $L$, and with envelope fitting manually turned off, $\langle T_\text{eff,ray}\rangle$ and $L_\text{ray}$. We calculate the luminosity $L_\text{3D,ph}$ that would be measured from a star with the same radius as the \mesa model $R_\text{1D}$ and the effective temperature from our envelope fitting method $\langle T_\text{eff}\rangle$. We provide the average surface particle optical depth $\langle\tau_{i,s}\rangle$ and average relative fractional change in specific entropy from initialization to relaxed $\langle\delta \left(s_{i,s}/k_bN_A\right)\rangle$ where $k_b$ is Boltzmann's constant and $N_A$ is Avogadro's number. All temperatures are given in Kelvins and all luminosities in $L_\odot$.}
	\label{tab:starsmasher_models}
    \begin{tabular}{llllllllll}
        \hline
        Model &
        $T_\text{eff,1D}$ &
        $\langle T_\text{eff}\rangle$ &
        $\langle T_\text{eff,ray}\rangle$ &
        $L_\text{1D}$ &
        $L$ &
        $L_\text{ray}$ &
        $L_\text{3D, ph}$ &
        $\langle\tau_{i,s}\rangle$ &
        $\langle\delta \left(s_{i,s}/k_bN_A\right)\rangle$ \\
        \hline
              \mesa & $ 5316$ &         &         & $ 6.69$ &         &         &         &         &          \\
\modelone   &         & $ 5766$ & $  1.61 \times 10^{5}$ &         & $ 10.5$ & $  6.82 \times 10^{6}$ & $ 9.27$ & $  6.93 \times 10^{9}$ & $ 2.76 \times 10^{-4}$  \\
\\
      \mesa & $ 4973$ &         &         & $ 7.59$ &         &         &         &         &          \\
\modeltwo   &         & $ 5151$ & $  1.27 \times 10^{5}$ &         & $ 9.76$ & $  3.85 \times 10^{6}$ & $ 8.73$ & $ 2.64 \times 10^{10}$ & $ 1.19 \times 10^{-4}$  \\
\modelthree &         & $ 5256$ & $  1.18 \times 10^{5}$ &         & $ 10.7$ & $  2.94 \times 10^{6}$ & $ 9.46$ & $ 1.31 \times 10^{10}$ & $ 1.55 \times 10^{-4}$  \\
\modelfour  &         & $ 5282$ & $  1.15 \times 10^{5}$ &         & $ 10.9$ & $  2.62 \times 10^{6}$ & $ 9.65$ & $  9.49 \times 10^{9}$ & $ 1.24 \times 10^{-4}$  \\
\\
      \mesa & $ 4577$ &         &         & $ 39.4$ &         &         &         &         &          \\
\modelfive  &         & $ 4665$ & $  5.25 \times 10^{4}$ &         & $ 47.1$ & $  8.26 \times 10^{5}$ & $ 42.5$ & $  4.82 \times 10^{9}$ & $ 6.90 \times 10^{-5}$  \\
        \hline
    \end{tabular}
\end{table*}

\begin{figure*}
	\centering
	\includegraphics[width=\textwidth]{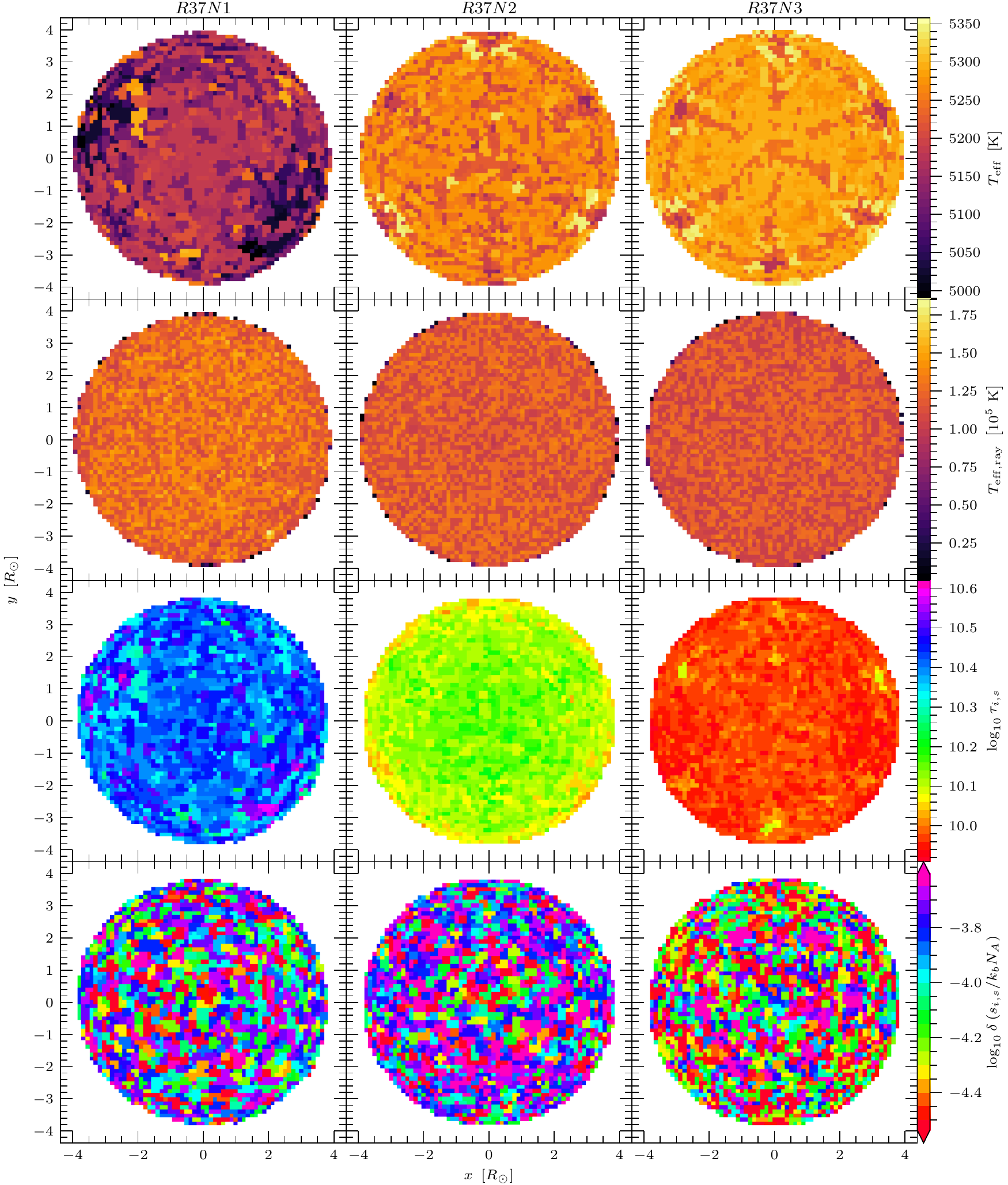}
    \caption{We show for our relaxed 3D stellar models \modeltwo (left), \modelthree (center), and \modelfour (right) the effective temperatures as calculated by our envelope fitting method only (top row) and ray tracing only (top-middle row) and surface particle optical depths $\tau_{i,s}$ (bottom-middle row) and the relative fractional change in specific entropy $\delta \left(s_{i,s}/k_bN_A\right)$ from initialization to relaxed (bottom row), where $k_b$ is Boltzmann's constant and $N_A$ is Avogadro's number. The grid has $63\times63$ cells and we trace each ray to $\tau=20$ where for each step $0<\Delta\tau\leq 0.1$.}
    \label{fig:teffs_and_taus}
\end{figure*}

\subsection{Single stars}\label{sec:single_stars}%
We obtain the luminosity for each of our 3D models using the adopted value of $\tau_\text{th}=20$, and requiring the convergence for the luminosity to be within 1\,per~cent.
The required precision was achieved using $63\times63$ grid cells.
We provide the values of the average effective temperatures $\langle T_\text{eff}\rangle$ and the corresponding total luminosities $L$ as measured by the observer for all our relaxed 3D models in \reftab{tab:starsmasher_models}.

The average surface particle optical depth $\langle\tau_{i,s}\rangle \gtrsim 10^9$. Due to this, effectively, for all our relaxed 3D models, $\langle T_\text{eff}\rangle$ and $L$ were obtained using only envelope fitting (case \ref{envfit_on}).
We find that $\langle T_\text{eff}\rangle$ is no more than 9\,per~cent greater than the effective temperatures of the 1D models $T_\text{eff,1D}$ (up to 451\,K) and $L$ is no more than 57\,per~cent greater than the luminosities of the 1D models $L_\text{1D}$.

We compare our standard method to that of ray tracing only, by manually turning off envelope fitting.
To clarify, we use the same value of $\tau_\text{th}=20$ as for our standard method, but for ray tracing it means that the integration into the surface particles stops at $\tau_\text{th}$.
We find for all our relaxed 3D models that $\langle T_\text{eff,ray}\rangle$ is from 11 to 30 times greater than $T_\text{eff,1D}$.
Accordingly, $L_\text{ray}$ is up to $10^6$ times larger than $L_\text{1D}$.
Thus, in using our envelope fitting method over ray tracing, we find an enormous improvement in accuracy for both temperature and luminosity of the relaxed models, providing us with the ability to obtain the light curve starting from before the interaction for the first time.

Using ray tracing, achieving a level of accuracy approximately similar to that of our envelope fitting method would require us to use a prohibitively large number of particles $N$.
In 1D stellar models, by definition, $T_\text{eff}\equiv T_\text{ph}$, where $T_\text{ph}$ is the temperature of the photosphere defined to be at the optical depth $\tau=2/3$.
The radius of our 1D models $R_\text{1D}$ is the location of the photosphere as well.
The surface particles in our 3D models would need to have temperatures of approximately $T_\text{ph}$ and thus be located near radius $R_\text{1D}$ and have a kernel size of $R-R_\text{1D}$ or smaller, where $R$ is the radius at $\tau=0$ in the 1D model.
In this way, as long as the SPH calculations can be trusted, the value of $T_\text{ph}$ can be known everywhere at the surface of the 3D model.
On initialization of a 3D model we hold the number density constant, which can be written as $3N/4\pi R_\text{1D} ^3$ and $3N_{nb,i}/4\pi (R-R_\text{1D} )^3$, and thus to construct our 3D model we need a number of particles
\begin{align}
    N = \left(\frac{R_\text{1D} }{R-R_\text{1D} }\right)^3 N_{nb,i}\ .
\end{align}
For our 1D model we used to generate \modeltwo, \modelthree, and \modelfour, we find $R-R_\text{1D} \approx 10^{-3}$\,$R_\odot$ by using \refeq{eq:optical_depth} with $\tau=2/3$ and a constant density and mean opacity equal to that of the 1D model at $R_\text{1D} \approx3.72$\,$R_\odot$, $\rho\simeq 8.3\times10^{-8}$\,g\,cm$^{-3}$ and $\bar{\kappa}\simeq0.12$\,cm$^2$\,g$^{-1}$.
Even in the minimal case of $N_{nb,i}=1$, which corresponds with grid-based methods, we must have $N \simeq 5\times10^{10}$ and for our 1D model we used to create \modelfour, $N \simeq 4\times10^9$.
These estimates represent the lower bound, as we do not account for the increase in particle smoothing length near the simulation boundary and $N_{nb,i}\gg1$ would be necessary for accurate SPH calculations.
The lower bound of $\sim10^{10}$ particles is close to the maximum ever achieved so far by any particle simulation method (see, e.g. the Millennium run \citealt{Springel_2005}, \texttt{VPIC} \citealt{2008PhPl...15e5703B}, \texttt{HACC} \citealt{2019arXiv190411970H}, and \texttt{Bonsai} \citealt{2014hpcn.conf...54B}).

We identify the error in our $L$ values as coming from the uncertainty in detecting our 3D models' radii, the uncertainty in our SPH simulations, and the uncertainty in our envelope fitting method.
All our relaxed 3D models have envelope radii larger than their 1D models if we let the radii be $R_{2h}$, but smaller if we let the radii be $R_\text{SPH}$ (see \reftab{tab:rel_models}).
This uncertainty in radius also appears in other SPH stellar models \citep[see, e.g.][]{2012ApJ...744...52P, 2017A&A...599A...5O, 2017MNRAS.464.4028I, 2019arXiv190709062J,2019MNRAS.484..631R}.
Our method to obtain luminosity is directly linked to integrating over the area that has non-zero column density along the rays.
As a result, the area contributing to the luminosity is bounded by $R_{2h}$, which is known to exceed the photospheric radius $R_\text{1D}$.
To investigate the effect the radius uncertainty has on $L$, we calculate for each of our 3D models the luminosity of a star that has a radius $R_\text{1D}$ and an effective temperature equal to $\langle T_\text{eff}\rangle$:
\begin{equation}
    L_\text{3D, ph} = 4\pi\sigma_\text{SB}R_\text{1D} ^2\langle T_\text{eff}\rangle^4\ .
\end{equation}
This luminosity exceeds the 1D value by $\sim$8--38\,per~cent.

There are two primary sources of the differences between 3D values and 1D values, gravity and entropy.
By default, we use the value of gravity at the location of the core of each SPH particle.
However, the `surface' of such a particle is located further away from the star's center and has a smaller surface gravity.
We tested what would be the effective temperature of our 3D star if the gravity is taken as in the 1D model.
While the `direct' value of $T_\text{eff}$ as obtained by our code for \modelfour is 5282\,K, feeding the post-processing routine with the value of the surface gravity as in the original 1D star, we obtain 5240\,K.
A similar value, 5230\,K is obtained when we calculated gravity in our code as if the surface of an SPH particle is located at one smoothing length further away from the center of the gravitational mass than the particle's center.
Further, during the relaxation process, SPH particles' entropies changed slightly compared to pre-relaxation.
If we take the same particles but adjust their pressure so that their entropy would match the entropy of the matter in the envelope of a 1D star, then the obtained value of surface effective temperature is 4950\,K.
Finally, if we correct both for surface gravity and entropy, we recover 4934\,K, which is very close to the value that the 1D star has.

We can also provide another way of estimating the error budget.
First, if we use values intrinsic to the initial 1D star, we can recover the effective temperatures, using the envelope fitting method, with a precision of $\pm 1$\,K.
However, if we change, for example, the  temperature of a particle by $\pm 0.01$ in its logarithmic value (or about 2.3\,per~cent in its linear value) while keeping the same pressure (so we change its entropy), then the obtained effective temperature would change similarly by about 2\,per~cent.
This deviation seems small; however, with the effective temperatures in the range of 5000\,K, it results in a 100\,K difference.
Uncertainty in the interpolated value of $\nabla$ by only 1\,per~cent (e.g., 0.2436 instead of 0.246) leads to 2.5\,per~cent deviation in the obtained effective temperature.
The effect of gravity is less linear: a change of its value by 5\,per~cent may cause almost no difference in effective temperature, but a variation by 10\,per~cent in gravity may lead to a 12\,per~cent difference because the envelope is shifting from convective to more radiative.
Our largest error is for the model \modelone, which is closest to the transition from radiative to convective.
Here, the difference between the SPH particle entropy and the initial 1D star entropy lead to different values of $\nabla$ and, consequently, to the largest deviations in the final value of the effective temperature ($8.5$\,per~cent).

We show in \reffig{fig:teffs_and_taus} our results for $T_\text{eff}$, $T_\text{eff,ray}$, $\log_{10}\tau_{i,s}$, and the relative fractional change in specific scaled entropy $\log_{10}\delta (s_{i,s}/k_bN_A)$ for the $63\times63$ grids we used in \reftab{tab:starsmasher_models} for \modeltwo, \modelthree, and \modelfour.
We find a hexagonally symmetric, `star-shaped' pattern in the distribution of $T_\text{eff}$ for \modelthree and \modelfour, which is likely a result of our initial placement of the SPH particles on an hcp lattice \citep[see Figure 16 of][for similar behavior in 2D SPH particles]{2012CoPhC.183.1641C}.
We do not observe a star-shaped pattern in any of our other 3D models, and the pattern is not visible in $T_\text{eff,ray}$ due to the numerical noise of the ray tracing method.

\begin{figure*}
	\centering
	\includegraphics{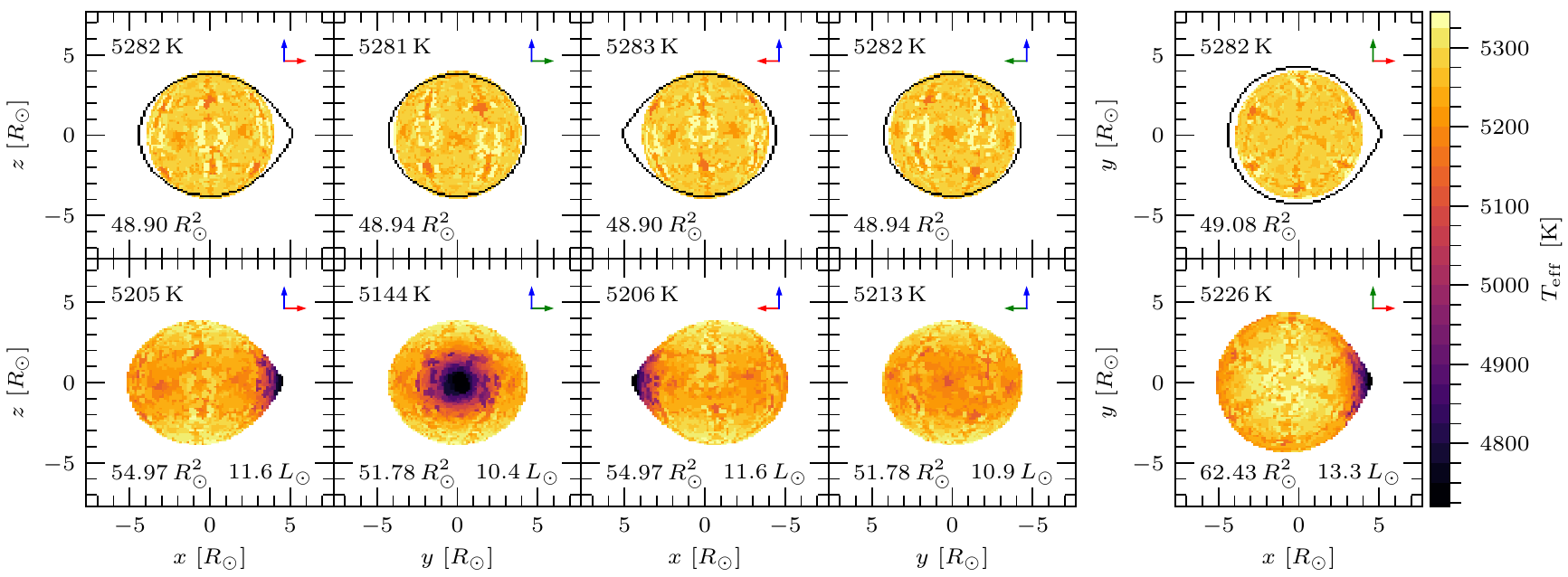}
    \caption{We show the spatial distribution of effective temperatures from 5 different viewing angles for our 3D model \modelfour (top panel) with a grid of $63\times63$ cells. We bring the same star to RLOF (bottom panel) in a binary with a $0.16$\,$M_\odot$ point mass companion (not visible). In the top-right corner of each panel we show the orientations of the $x$ axis (red), $y$ axis (green), and $z$ axis (blue). We calculate the average effective temperature $\langle T_\text{eff}\rangle$ (top left text), luminosity $L$ (bottom right text), and the sum of the areas of grid cells which have fluid somewhere along the associated ray (bottom left text). We compare the apparent shape of the RLOF star to that of \modelfour (black outline in the top panels) and find the envelope of the RLOF star to be oblate in comparison, which is likely due to the tidally-locked interaction. The luminosity of the single star shown in the top panel is always 10.9\,$L_\odot$ regardless of the angle.}
    \label{fig:RLOF_all_angles}
\end{figure*}

\subsection{Initial state in a binary setup}\label{sec:initial_state_in_a_binary_setup}%
In future work, we plan to obtain a light curve of a dynamically merging binary consisting of $M=1.52$\,$M_\odot$ models with a $0.16$\,$M_\odot$ secondary, similar to \cite{2014ApJ...786...39N}.
In this paper, we create the initial conditions for future calculations of a binary merger event.

We start by placing our relaxed single 3D model in a co-rotating frame with a $0.16$\,$M_\odot$ point mass secondary and scanning the binary equilibrium sequence \citep[see \refseclabel2.3 of][for details of this method]{2011ApJ...737...49L}.
Following that method, we first let \modelfour relax at an initial separation $a_i=30$\,$R_\odot$ from the $0.16$\,$M_\odot$ secondary for 100 dynamical times $t_\text{dyn}\equiv(R_\odot^3/GM_\odot)^{1/2}$.
We set the smoothing length of the point mass companion to be the same as that of the core particle in our donor star.
Then, we proceed with the scanning until the primary reaches Roche lobe overflow (hereafter called RLOF).

For the analyses, we choose a snapshot of the simulation when it is near RLOF, with orbital separation $a\simeq6.8$\,$R_\odot$ and period $P\simeq1.6$\,days.
We show in \reffig{fig:RLOF_all_angles} the visual-surface averaged value of $\langle T_\text{eff}\rangle$, and total $L$ that would be inferred by the observer from 5 different viewing angles.
The star's apparent shape is that of a `teardrop', as expected for stars near RLOF.
The distribution of surface temperatures has changed as compared to a single star, with the polar regions being hotter than the equatorial belt. 
A cool spot that is about 560\,K cooler than $\langle T_\text{eff}\rangle$ has formed on the side of the star facing the secondary.
Overall, from all the angles, the star is cooler than the single star model we started with by at most 137\,K, but the total inferred luminosity in each direction has not changed much.
We find that the effective visible area is, from all angles, larger than that of \modelfour due to both the filling of the Roche lobe and the envelope becoming somewhat oblate from the tidally-locked orbit.

We note that no limb darkening effect can be visible from any angle: all surface particles have $\tau_i \gg \tau_\text{th}=20$ and thus $T_\text{eff}$ is calculated for each ray using only our envelope fitting method, which is independent of ray angle.

%% file: Section5.tex
\section{Conclusions}\label{sec:conclusions}%
Our ultimate goal is to produce a synthetic light curve of a full CE evolution and compare it to observed light curves of transient events.
We expect the early-time morphology of such a synthetic light curve will be most affected by the radiative flux of the binary components and ejecta: such radiative flux is a strong function of the initial conditions.
In this paper, we present a method for creating 3D SPH stellar models whose physical properties closely match that of their initial 1D models, particularly in the upper portion of their envelopes, where we anticipate most of the dynamical ejecta will originate.
We also introduce a new method, which we call `envelope fitting', for obtaining the radiative flux of intrinsically optically thick SPH particles.
The total flux will be obtained as the flux emitted by the optically thick particles, attenuated by semi-opaque SPH particles.

Our envelope fitting method links an SPH particle's temperature and pressure to the effective temperature and surface pressure of a unique stellar envelope solution from a wide selection of envelope solutions in the parameter space of surface gravity.
It uses a realistic temperature gradient that is intrinsic to the envelope instead of the simplified temperature gradient usually employed (such as, for example, a stratified temperature gradient, e.g. \citealt{2017ApJS..229...36G,2017ApJ...850...59P,2021arXiv210205662B}). 

To test our method for creating our initial donor stars, we create several 3D SPH models of a possible V1309~Sco progenitor donor star for use in future CE evolution simulations.
We match the integrated total energy profiles of our 3D models to that of their initial 1D models to within 0.1\,per~cent in the outer envelope.
We accomplish this matching by varying an input parameter, \texttt{nnopt}, which controls the number of neighbors for each SPH particle.
For each of our created 3D stars, we obtain the luminosity and effective temperature and discuss the nature of the differences in each in comparison to our 1D models.
As an example of the initial conditions for our future mergers, we place our highest resolution star in a binary configuration with a point-mass companion and analyze the corresponding change in luminosity, effective temperature, and envelope shape.

Specifically, we find that:
\begin{itemize}
    \item For each 3D stellar model a single value of \texttt{nnopt} provides the best fit of the 3D star's integrated total energy profile to that of the initial 1D model. 
    For each stellar model, that number also depends on the adopted resolution.
    
    \item The luminosities that we obtain in 3D while using our ray tracing technique are very inaccurate when compared to the luminosities of the initial 1D models, exceeding such by more than a factor of $10^5$.
    The primary reason for this inaccuracy is that neither the SPH kernel gradient nor the adiabatic gradient adequately describe the realistic temperature gradient within optically thick surface SPH particles.
    The ray tracing technique therefore cannot help with modeling a light curve of the the initial stage of a stellar merger or a common envelope event.
    
    \item The use of stellar envelope solutions provides luminosities from our 3D models that match that of their 1D models to within 57\,per~cent, which corresponds to a difference in the observed magnitude of only up to $0.5$.
    We match the effective temperature, which is useful for spectral identification, to within a few per~cent.
    
    \item In a binary configuration, our 3D donor model is, depending on the viewing angle, cooler than its 1D model by no more than 137\,K (about 3\,per~cent) and its luminosity is, depending on the viewing angle, brighter by 22\,per~cent or dimmer by 5\,per~cent than our single unperturbed 3D star.
    The amplitude of the bolometric magnitude for the same star but as measured from different observer angles is $\sim0.3$, which is comparable to the pre-outburst magnitude oscillation of the V1309~Sco progenitor (see Figure 3 in \citealt{2011A&A...528A.114T}).
\end{itemize}

Our envelope fitting technique is not specific to SPH codes or Lagrangian methods and can be used in any context provided there exist unique solutions in the $P$--$T$ parameter space.
However, it has limitations.
The envelope fitting method links the SPH particle temperature (about 2--$6\times 10^5$\,K) with surface temperatures that are 100 times smaller.
Hence, a deviation on the order of 0.1\,per~cent in an SPH particle's central thermodynamic properties from 1D values at the same location intrinsically leads to about a 1\,per~cent spread in effective temperatures, and then to a few per~cent deviation in the apparent luminosity.

In the follow-up paper we plan to obtain the synthetic light curve of a merger event from a binary which we will create using all the steps described in this paper.
The initial conditions will likely strongly affect the outcomes of the full-fledged merger simulation.
For example, the overall energetics of the event and the outburst itself has been indicated to be influenced by the properties of $\lesssim0.05$\,$M_\odot$ ejecta  \citep{2013Sci...339..433I,2014ApJ...786...39N,2017ApJ...850...59P}.
To account for this, we have learned in this paper how to match the upper $0.1$\,$M_\odot$ of the donor's envelope with the integrated total energy profile to that of the initial 1D model.
The morphology of the light curve during the pre-outburst, which is the phase leading up to plunge-in, has to be modelled from the stage when the almost-unperturbed donor is observed.
Ability to model low-luminosity objects like binary components may allow us to recreate the growth of the light curve and see if it will be affected by internal shocks between merging spiral arms in the outflow as, for example, argued by \citet{2016MNRAS.455.4351P}, or if it will be rather affected by the expansion of the ejecta with a cooling front propagating into it, as argued by \citet{2013Sci...339..433I}.

We remind the reader that the method we describe in this paper for obtaining radiative fluxes is a post-processing method.
For the stationary stars that we consider, the thermal timescale on which the energy can be lost by radiation from the surface layers exceeds the characteristic timescale of the simulation (several dynamical timescales) by several orders of magnitude.
For a merger event, where the energy radiated away becomes comparable to the initially stored thermal energy and to the additional energy from the shocks, radiative cooling would need to be taken into account.
The post-processing method that we have developed finds the direction-dependent observed radiative flux, not the complete radiative `sink' at each specific place.
Accounting for radiative losses during a dynamical simulation should be done with a different approach, such as, for example, introducing a cooling term, as described in \citet{2015MNRAS.447...25L}.

The pre-outburst phase is relatively difficult to identify in photometric datasets, as it occurs over just a few thousand binary orbits (as suggested by the V1309~Sco OGLE data), which is only a few years for short-period binaries.
Pre-outburst CE binaries with small inclinations can have small binary variability  amplitudes, making them challenging to discover in archival data, and those that are viewed close to edge-on can quickly be obscured by dust after the onset of $L_2$ outflow.
So far, systematic transient surveys like the Zwicky Transient Facility (ZTF) have been searching for signatures of CE evolution by identifying pulsations in luminosity and temperature, but similar pulsations can also be produced by other objects such as luminous blue variable stars.
Nonetheless, we hope that systematic transient surveys like ZTF will provide us with more systems similar to V1309~Sco in the future, such that the initial stage of common envelope events may be elucidated.

%% file: Acknowledgements.tex
\section*{Acknowledgements}
We would like the thank the referee for comments that helped to improve the manuscript.
R.H. acknowledges Kenny X. Van, Zhuo Chen, Margaret E. Ridder, and Leon Olifer for their helpful discussions.
N.I. acknowledges support from CRC program and funding from NSERC Discovery under Grant No. NSERC RGPIN-2019-04277.
The computations were enabled by support provided by WestGrid (www.westgrid.ca) and Compute Canada Calcul Canada (www.computecanada.ca).

%% file: main.bbl
\begin{thebibliography}{}
\makeatletter
\relax
\def\mn@urlcharsother{\let\do\@makeother \do\$\do\&\do\#\do\^\do\_\do\%\do\~}
\def\mn@doi{\begingroup\mn@urlcharsother \@ifnextchar [ {\mn@doi@}
  {\mn@doi@[]}}
\def\mn@doi@[#1]#2{\def\@tempa{#1}\ifx\@tempa\@empty \href
  {http://dx.doi.org/#2} {doi:#2}\else \href {http://dx.doi.org/#2} {#1}\fi
  \endgroup}
\def\mn@eprint#1#2{\mn@eprint@#1:#2::\@nil}
\def\mn@eprint@arXiv#1{\href {http://arxiv.org/abs/#1} {{\tt arXiv:#1}}}
\def\mn@eprint@dblp#1{\href {http://dblp.uni-trier.de/rec/bibtex/#1.xml}
  {dblp:#1}}
\def\mn@eprint@#1:#2:#3:#4\@nil{\def\@tempa {#1}\def\@tempb {#2}\def\@tempc
  {#3}\ifx \@tempc \@empty \let \@tempc \@tempb \let \@tempb \@tempa \fi \ifx
  \@tempb \@empty \def\@tempb {arXiv}\fi \@ifundefined
  {mn@eprint@\@tempb}{\@tempb:\@tempc}{\expandafter \expandafter \csname
  mn@eprint@\@tempb\endcsname \expandafter{\@tempc}}}

\bibitem[\protect\citeauthoryear{{Altay} \& {Theuns}}{{Altay} \&
  {Theuns}}{2013}]{2013MNRAS.434..748A}
{Altay} G.,  {Theuns} T.,  2013, \mn@doi [\mnras] {10.1093/mnras/stt1067},
  \href {https://ui.adsabs.harvard.edu/\#abs/2013MNRAS.434..748A} {434, 748}

\bibitem[\protect\citeauthoryear{{Altay}, {Croft}  \& {Pelupessy}}{{Altay}
  et~al.}{2008}]{2008MNRAS.386.1931A}
{Altay} G.,  {Croft} R. A.~C.,   {Pelupessy} I.,  2008, \mn@doi [\mnras]
  {10.1111/j.1365-2966.2008.13212.x}, \href
  {https://ui.adsabs.harvard.edu/\#abs/2008MNRAS.386.1931A} {386, 1931}

\bibitem[\protect\citeauthoryear{{B{\'e}dorf}, {Gaburov}, {Fujii}, {Nitadori},
  {Ishiyama}  \& {Portegies Zwart}}{{B{\'e}dorf}
  et~al.}{2014}]{2014hpcn.conf...54B}
{B{\'e}dorf} J.,  {Gaburov} E.,  {Fujii} M.~S.,  {Nitadori} K.,  {Ishiyama} T.,
    {Portegies Zwart} S.,  2014, in Proceedings of the International Conference
  for High Performance Computing. pp 54--65 (\mn@eprint {arXiv} {1412.0659}),
  \mn@doi{10.1109/SC.2014.10}

\bibitem[\protect\citeauthoryear{{Belczynski}, {Holz}, {Bulik}  \&
  {O'Shaughnessy}}{{Belczynski} et~al.}{2016}]{2016Natur.534..512B}
{Belczynski} K.,  {Holz} D.~E.,  {Bulik} T.,   {O'Shaughnessy} R.,  2016,
  \mn@doi [\nat] {10.1038/nature18322}, \href
  {https://ui.adsabs.harvard.edu/abs/2016Natur.534..512B} {534, 512}

\bibitem[\protect\citeauthoryear{{Blagorodnova} et~al.,}{{Blagorodnova}
  et~al.}{2021}]{2021arXiv210205662B}
{Blagorodnova} N.,  et~al., 2021, arXiv e-prints, \href
  {https://ui.adsabs.harvard.edu/abs/2021arXiv210205662B} {p. arXiv:2102.05662}

\bibitem[\protect\citeauthoryear{{Bowers}, {Albright}, {Yin}, {Bergen}  \&
  {Kwan}}{{Bowers} et~al.}{2008}]{2008PhPl...15e5703B}
{Bowers} K.~J.,  {Albright} B.~J.,  {Yin} L.,  {Bergen} B.,   {Kwan} T.~J.~T.,
  2008, \mn@doi [Physics of Plasmas] {10.1063/1.2840133}, \href
  {https://ui.adsabs.harvard.edu/abs/2008PhPl...15e5703B} {15, 055703}

\bibitem[\protect\citeauthoryear{{Britavskiy} et~al.,}{{Britavskiy}
  et~al.}{2019}]{2019A&A...624A.128B}
{Britavskiy} N.,  et~al., 2019, \mn@doi [\aap] {10.1051/0004-6361/201834564},
  \href {https://ui.adsabs.harvard.edu/abs/2019A&A...624A.128B} {624, A128}

\bibitem[\protect\citeauthoryear{{Bujarrabal}, {Garc{\'\i}a-Segura}, {Morris},
  {Soker}  \& {Terzian}}{{Bujarrabal} et~al.}{2000}]{2000ASPC..199..201B}
{Bujarrabal} V.,  {Garc{\'\i}a-Segura} G.,  {Morris} M.,  {Soker} N.,
  {Terzian} Y.,  2000, in {Kastner} J.~H.,  {Soker} N.,   {Rappaport} S.,  eds,
   Astronomical Society of the Pacific Conference Series Vol. 199, Asymmetrical
  Planetary Nebulae II: From Origins to Microstructures. p.~201

\bibitem[\protect\citeauthoryear{{Colagrossi}, {Bouscasse}, {Antuono}  \&
  {Marrone}}{{Colagrossi} et~al.}{2012}]{2012CoPhC.183.1641C}
{Colagrossi} A.,  {Bouscasse} B.,  {Antuono} M.,   {Marrone} S.,  2012, \mn@doi
  [Computer Physics Communications] {10.1016/j.cpc.2012.02.032}, \href
  {https://ui.adsabs.harvard.edu/abs/2012CoPhC.183.1641C} {183, 1641}

\bibitem[\protect\citeauthoryear{{Dehnen} \& {Aly}}{{Dehnen} \&
  {Aly}}{2012}]{2012MNRAS.425.1068D}
{Dehnen} W.,  {Aly} H.,  2012, \mn@doi [\mnras]
  {10.1111/j.1365-2966.2012.21439.x}, \href
  {http://adsabs.harvard.edu/abs/2012MNRAS.425.1068D} {425, 1068}

\bibitem[\protect\citeauthoryear{{Dominik}, {Belczynski}, {Fryer}, {Holz},
  {Berti}, {Bulik}, {Mandel}  \& {O'Shaughnessy}}{{Dominik}
  et~al.}{2012}]{2012ApJ...759...52D}
{Dominik} M.,  {Belczynski} K.,  {Fryer} C.,  {Holz} D.~E.,  {Berti} E.,
  {Bulik} T.,  {Mandel} I.,   {O'Shaughnessy} R.,  2012, \mn@doi [\apj]
  {10.1088/0004-637X/759/1/52}, \href
  {https://ui.adsabs.harvard.edu/abs/2012ApJ...759...52D} {759, 52}

\bibitem[\protect\citeauthoryear{{Ferguson}, {Alexander}, {Allard}, {Barman},
  {Bodnarik}, {Hauschildt}, {Heffner-Wong}  \& {Tamanai}}{{Ferguson}
  et~al.}{2005}]{Ferguson2005}
{Ferguson} J.~W.,  {Alexander} D.~R.,  {Allard} F.,  {Barman} T.,  {Bodnarik}
  J.~G.,  {Hauschildt} P.~H.,  {Heffner-Wong} A.,   {Tamanai} A.,  2005,
  \mn@doi [\apj] {10.1086/428642}, \href
  {https://ui.adsabs.harvard.edu/abs/2005ApJ...623..585F} {623, 585}

\bibitem[\protect\citeauthoryear{{Ferreira}, {Saito}, {Minniti}, {Navarro},
  {Ramos}, {Smith}  \& {Lucas}}{{Ferreira} et~al.}{2019}]{2019MNRAS.486.1220F}
{Ferreira} T.,  {Saito} R.~K.,  {Minniti} D.,  {Navarro} M.~G.,  {Ramos} R.~C.,
   {Smith} L.,   {Lucas} P.~W.,  2019, \mn@doi [\mnras] {10.1093/mnras/stz878},
  \href {https://ui.adsabs.harvard.edu/abs/2019MNRAS.486.1220F} {486, 1220}

\bibitem[\protect\citeauthoryear{{Gaburov}, {Lombardi}  \& {Portegies
  Zwart}}{{Gaburov} et~al.}{2010}]{2010MNRAS.402..105G}
{Gaburov} E.,  {Lombardi} Jr. J.~C.,   {Portegies Zwart} S.,  2010, \mn@doi
  [\mnras] {10.1111/j.1365-2966.2009.15900.x}, \href
  {http://adsabs.harvard.edu/abs/2010MNRAS.402..105G} {402, 105}

\bibitem[\protect\citeauthoryear{{Galaviz}, {De Marco}, {Passy}, {Staff}  \&
  {Iaconi}}{{Galaviz} et~al.}{2017}]{2017ApJS..229...36G}
{Galaviz} P.,  {De Marco} O.,  {Passy} J.-C.,  {Staff} J.~E.,   {Iaconi} R.,
  2017, \mn@doi [\apjs] {10.3847/1538-4365/aa64e1}, \href
  {https://ui.adsabs.harvard.edu/abs/2017ApJS..229...36G} {229, 36}

\bibitem[\protect\citeauthoryear{{Han}, {Podsiadlowski}, {Maxted}, {Marsh}  \&
  {Ivanova}}{{Han} et~al.}{2002}]{2002MNRAS.336..449H}
{Han} Z.,  {Podsiadlowski} P.,  {Maxted} P.~F.~L.,  {Marsh} T.~R.,   {Ivanova}
  N.,  2002, \mn@doi [\mnras] {10.1046/j.1365-8711.2002.05752.x}, \href
  {https://ui.adsabs.harvard.edu/abs/2002MNRAS.336..449H} {336, 449}

\bibitem[\protect\citeauthoryear{{Heitmann} et~al.,}{{Heitmann}
  et~al.}{2019}]{2019arXiv190411970H}
{Heitmann} K.,  et~al., 2019, arXiv e-prints, \href
  {https://ui.adsabs.harvard.edu/abs/2019arXiv190411970H} {p. arXiv:1904.11970}

\bibitem[\protect\citeauthoryear{{Howitt}, {Stevenson}, {Vigna-G{\'o}mez},
  {Justham}, {Ivanova}, {Woods}, {Neijssel}  \& {Mandel}}{{Howitt}
  et~al.}{2020}]{2020MNRAS.492.3229H}
{Howitt} G.,  {Stevenson} S.,  {Vigna-G{\'o}mez} A.,  {Justham} S.,  {Ivanova}
  N.,  {Woods} T.~E.,  {Neijssel} C.~J.,   {Mandel} I.,  2020, \mn@doi [\mnras]
  {10.1093/mnras/stz3542}, \href
  {https://ui.adsabs.harvard.edu/abs/2020MNRAS.492.3229H} {492, 3229}

\bibitem[\protect\citeauthoryear{{Iaconi}, {Reichardt}, {Staff}, {De Marco},
  {Passy}, {Price}, {Wurster}  \& {Herwig}}{{Iaconi}
  et~al.}{2017}]{2017MNRAS.464.4028I}
{Iaconi} R.,  {Reichardt} T.,  {Staff} J.,  {De Marco} O.,  {Passy} J.-C.,
  {Price} D.,  {Wurster} J.,   {Herwig} F.,  2017, \mn@doi [\mnras]
  {10.1093/mnras/stw2377}, \href
  {https://ui.adsabs.harvard.edu/abs/2017MNRAS.464.4028I} {464, 4028}

\bibitem[\protect\citeauthoryear{{Iben} \& {Livio}}{{Iben} \&
  {Livio}}{1993}]{1993PASP..105.1373I}
{Iben} Icko J.,  {Livio} M.,  1993, \mn@doi [\pasp] {10.1086/133321}, \href
  {https://ui.adsabs.harvard.edu/abs/1993PASP..105.1373I} {105, 1373}

\bibitem[\protect\citeauthoryear{{Ivanova}, {Justham}, {Avendano Nandez}  \&
  {Lombardi}}{{Ivanova} et~al.}{2013}]{2013Sci...339..433I}
{Ivanova} N.,  {Justham} S.,  {Avendano Nandez} J.~L.,   {Lombardi} J.~C.,
  2013, \mn@doi [Science] {10.1126/science.1225540}, \href
  {https://ui.adsabs.harvard.edu/abs/2013Sci...339..433I} {339, 433}

\bibitem[\protect\citeauthoryear{Ivanova, Justham  \& Ricker}{Ivanova
  et~al.}{2020}]{comenv_book_2020}
Ivanova N.,  Justham S.,   Ricker P.,  2020, Common Envelope Evolution.
2514-3433, IOP Publishing, \mn@doi{10.1088/2514-3433/abb6f0}, \url
  {http://dx.doi.org/10.1088/2514-3433/abb6f0}

\bibitem[\protect\citeauthoryear{{Jones} \& {Boffin}}{{Jones} \&
  {Boffin}}{2017}]{2017NatAs...1E.117J}
{Jones} D.,  {Boffin} H. M.~J.,  2017, \mn@doi [Nature Astronomy]
  {10.1038/s41550-017-0117}, \href
  {https://ui.adsabs.harvard.edu/abs/2017NatAs...1E.117J} {1, 0117}

\bibitem[\protect\citeauthoryear{{Joyce}, {Lairmore}, {Price}, {Reichardt}  \&
  {Mohamed}}{{Joyce} et~al.}{2019}]{2019arXiv190709062J}
{Joyce} M.,  {Lairmore} L.,  {Price} D.~J.,  {Reichardt} T.,   {Mohamed} S.,
  2019, arXiv e-prints, \href
  {https://ui.adsabs.harvard.edu/abs/2019arXiv190709062J} {p. arXiv:1907.09062}

\bibitem[\protect\citeauthoryear{{Kami{\'n}ski}, {Steffen}, {Tylenda}, {Young},
  {Patel}  \& {Menten}}{{Kami{\'n}ski} et~al.}{2018}]{2018A&A...617A.129K}
{Kami{\'n}ski} T.,  {Steffen} W.,  {Tylenda} R.,  {Young} K.~H.,  {Patel}
  N.~A.,   {Menten} K.~M.,  2018, \mn@doi [\aap] {10.1051/0004-6361/201833165},
  \href {https://ui.adsabs.harvard.edu/abs/2018A&A...617A.129K} {617, A129}

\bibitem[\protect\citeauthoryear{{Kami{\'n}ski}, {Steffen}, {Bujarrabal},
  {Tylenda}, {Menten}  \& {Hajduk}}{{Kami{\'n}ski}
  et~al.}{2021}]{2021A&A...646A...1K}
{Kami{\'n}ski} T.,  {Steffen} W.,  {Bujarrabal} V.,  {Tylenda} R.,  {Menten}
  K.~M.,   {Hajduk} M.,  2021, \mn@doi [\aap] {10.1051/0004-6361/202039634},
  \href {https://ui.adsabs.harvard.edu/abs/2021A&A...646A...1K} {646, A1}

\bibitem[\protect\citeauthoryear{{Kasliwal}}{{Kasliwal}}{2012}]{2012PASA...29..482K}
{Kasliwal} M.~M.,  2012, \mn@doi [\pasa] {10.1071/AS11061}, \href
  {https://ui.adsabs.harvard.edu/abs/2012PASA...29..482K} {29, 482}

\bibitem[\protect\citeauthoryear{{Kippenhahn}, {Weigert}  \&
  {Weiss}}{{Kippenhahn} et~al.}{2012}]{2012sse..book.....K}
{Kippenhahn} R.,  {Weigert} A.,   {Weiss} A.,  2012, {Stellar Structure and
  Evolution}.
Springer-Verlag Berlin Heidelberg, \mn@doi{10.1007/978-3-642-30304-3}

\bibitem[\protect\citeauthoryear{{Kittel}}{{Kittel}}{1976}]{1976itss.book.....K}
{Kittel} C.,  1976, {Introduction to solid state physics}.
Wiley

\bibitem[\protect\citeauthoryear{{Klencki}, {Nelemans}, {Istrate}  \&
  {Chruslinska}}{{Klencki} et~al.}{2021}]{2021A&A...645A..54K}
{Klencki} J.,  {Nelemans} G.,  {Istrate} A.~G.,   {Chruslinska} M.,  2021,
  \mn@doi [\aap] {10.1051/0004-6361/202038707}, \href
  {https://ui.adsabs.harvard.edu/abs/2021A&A...645A..54K} {645, A54}

\bibitem[\protect\citeauthoryear{{Kochanek}, {Adams}  \&
  {Belczynski}}{{Kochanek} et~al.}{2014}]{2014MNRAS.443.1319K}
{Kochanek} C.~S.,  {Adams} S.~M.,   {Belczynski} K.,  2014, \mn@doi [\mnras]
  {10.1093/mnras/stu1226}, \href
  {https://ui.adsabs.harvard.edu/abs/2014MNRAS.443.1319K} {443, 1319}

\bibitem[\protect\citeauthoryear{{Lipunov} et~al.,}{{Lipunov}
  et~al.}{2017}]{2017MNRAS.470.2339L}
{Lipunov} V.~M.,  et~al., 2017, \mn@doi [\mnras] {10.1093/mnras/stx1107}, \href
  {https://ui.adsabs.harvard.edu/abs/2017MNRAS.470.2339L} {470, 2339}

\bibitem[\protect\citeauthoryear{{Lombardi}, {Proulx}, {Dooley}, {Theriault},
  {Ivanova}  \& {Rasio}}{{Lombardi} et~al.}{2006}]{2006ApJ...640..441L}
{Lombardi} J.~C. J.,  {Proulx} Z.~F.,  {Dooley} K.~L.,  {Theriault} E.~M.,
  {Ivanova} N.,   {Rasio} F.~A.,  2006, \mn@doi [\apj] {10.1086/499938}, \href
  {https://ui.adsabs.harvard.edu/#abs/2006ApJ...640..441L} {640, 441}

\bibitem[\protect\citeauthoryear{{Lombardi}, {Holtzman}, {Dooley}, {Gearity},
  {Kalogera}  \& {Rasio}}{{Lombardi} et~al.}{2011}]{2011ApJ...737...49L}
{Lombardi} J.~C. J.,  {Holtzman} W.,  {Dooley} K.~L.,  {Gearity} K.,
  {Kalogera} V.,   {Rasio} F.~A.,  2011, \mn@doi [\apj]
  {10.1088/0004-637X/737/2/49}, \href
  {https://ui.adsabs.harvard.edu/abs/2011ApJ...737...49L} {737, 49}

\bibitem[\protect\citeauthoryear{{Lombardi}, {McInally}  \& {Faber}}{{Lombardi}
  et~al.}{2015}]{2015MNRAS.447...25L}
{Lombardi} J.~C.,  {McInally} W.~G.,   {Faber} J.~A.,  2015, \mn@doi [\mnras]
  {10.1093/mnras/stu2432}, \href
  {https://ui.adsabs.harvard.edu/abs/2015MNRAS.447...25L} {447, 25}

\bibitem[\protect\citeauthoryear{{McCrea}}{{McCrea}}{1964}]{1964MNRAS.128..147M}
{McCrea} W.~H.,  1964, \mn@doi [\mnras] {10.1093/mnras/128.2.147}, \href
  {https://ui.adsabs.harvard.edu/abs/1964MNRAS.128..147M} {128, 147}

\bibitem[\protect\citeauthoryear{{Metzger} \& {Pejcha}}{{Metzger} \&
  {Pejcha}}{2017}]{2017MNRAS.471.3200M}
{Metzger} B.~D.,  {Pejcha} O.,  2017, \mn@doi [\mnras] {10.1093/mnras/stx1768},
  \href {https://ui.adsabs.harvard.edu/abs/2017MNRAS.471.3200M} {471, 3200}

\bibitem[\protect\citeauthoryear{{Monaghan}}{{Monaghan}}{1992}]{1992ARA&A..30..543M}
{Monaghan} J.~J.,  1992, \mn@doi [\araa] {10.1146/annurev.aa.30.090192.002551},
  \href {http://adsabs.harvard.edu/abs/1992ARA%26A..30..543M} {30, 543}

\bibitem[\protect\citeauthoryear{{Mor{\'e}}}{{Mor{\'e}}}{1978}]{1978LNM...630..105M}
{Mor{\'e}} J.~J.,  1978, {The Levenberg-Marquardt algorithm: Implementation and
  theory}.
Springer Berlin Heidelberg, pp 105--116, \mn@doi{10.1007/BFb0067700}

\bibitem[\protect\citeauthoryear{{Nandez}, {Ivanova}  \& {Lombardi}}{{Nandez}
  et~al.}{2014}]{2014ApJ...786...39N}
{Nandez} J.~L.~A.,  {Ivanova} N.,   {Lombardi} J.~C. J.,  2014, \mn@doi [\apj]
  {10.1088/0004-637X/786/1/39}, \href
  {https://ui.adsabs.harvard.edu/abs/2014ApJ...786...39N} {786, 39}

\bibitem[\protect\citeauthoryear{{Nandez}, {Ivanova}  \& {Lombardi}}{{Nandez}
  et~al.}{2015}]{2015MNRAS.450L..39N}
{Nandez} J.~L.~A.,  {Ivanova} N.,   {Lombardi} J.~C.~J.,  2015, \mn@doi
  [\mnras] {10.1093/mnrasl/slv043}, \href
  {https://ui.adsabs.harvard.edu/\#abs/2015MNRAS.450L..39N} {450, L39}

\bibitem[\protect\citeauthoryear{{Natale}, {Popescu}, {Tuffs}, {Debattista},
  {Fischera}  \& {Grootes}}{{Natale} et~al.}{2015}]{2015MNRAS.449..243N}
{Natale} G.,  {Popescu} C.~C.,  {Tuffs} R.~J.,  {Debattista} V.~P.,  {Fischera}
  J.,   {Grootes} M.~W.,  2015, \mn@doi [\mnras] {10.1093/mnras/stv286}, \href
  {https://ui.adsabs.harvard.edu/\#abs/2015MNRAS.449..243N} {449, 243}

\bibitem[\protect\citeauthoryear{{Ohlmann}, {R{\"o}pke}, {Pakmor}  \&
  {Springel}}{{Ohlmann} et~al.}{2017}]{2017A&A...599A...5O}
{Ohlmann} S.~T.,  {R{\"o}pke} F.~K.,  {Pakmor} R.,   {Springel} V.,  2017,
  \mn@doi [\aap] {10.1051/0004-6361/201629692}, \href
  {https://ui.adsabs.harvard.edu/abs/2017A&A...599A...5O} {599, A5}

\bibitem[\protect\citeauthoryear{{Paczynski}}{{Paczynski}}{1976}]{1976IAUS...73...75P}
{Paczynski} B.,  1976, in {Eggleton} P.,  {Mitton} S.,   {Whelan} J.,  eds,
  IAU Symposium Vol. 73, Structure and Evolution of Close Binary Systems. p.~75

\bibitem[\protect\citeauthoryear{{Passy} et~al.,}{{Passy}
  et~al.}{2012}]{2012ApJ...744...52P}
{Passy} J.-C.,  et~al., 2012, \mn@doi [\apj] {10.1088/0004-637X/744/1/52},
  \href {https://ui.adsabs.harvard.edu/abs/2012ApJ...744...52P} {744, 52}

\bibitem[\protect\citeauthoryear{{Pastorello} et~al.,}{{Pastorello}
  et~al.}{2020}]{2020arXiv201110590P}
{Pastorello} A.,  et~al., 2020, arXiv e-prints, \href
  {https://ui.adsabs.harvard.edu/abs/2020arXiv201110590P} {p. arXiv:2011.10590}

\bibitem[\protect\citeauthoryear{{Pastorello} et~al.,}{{Pastorello}
  et~al.}{2021}]{2021A&A...646A.119P}
{Pastorello} A.,  et~al., 2021, \mn@doi [\aap] {10.1051/0004-6361/202039952},
  \href {https://ui.adsabs.harvard.edu/abs/2021A&A...646A.119P} {646, A119}

\bibitem[\protect\citeauthoryear{{Paxton}, {Bildsten}, {Dotter}, {Herwig},
  {Lesaffre}  \& {Timmes}}{{Paxton} et~al.}{2011}]{Paxton2011}
{Paxton} B.,  {Bildsten} L.,  {Dotter} A.,  {Herwig} F.,  {Lesaffre} P.,
  {Timmes} F.,  2011, \mn@doi [\apjs] {10.1088/0067-0049/192/1/3}, \href
  {https://ui.adsabs.harvard.edu/abs/2011ApJS..192....3P} {192, 3}

\bibitem[\protect\citeauthoryear{{Paxton} et~al.,}{{Paxton}
  et~al.}{2013}]{Paxton2013}
{Paxton} B.,  et~al., 2013, \mn@doi [\apjs] {10.1088/0067-0049/208/1/4}, \href
  {https://ui.adsabs.harvard.edu/abs/2013ApJS..208....4P} {208, 4}

\bibitem[\protect\citeauthoryear{{Paxton} et~al.,}{{Paxton}
  et~al.}{2015}]{Paxton2015}
{Paxton} B.,  et~al., 2015, \mn@doi [\apjs] {10.1088/0067-0049/220/1/15}, \href
  {https://ui.adsabs.harvard.edu/abs/2015ApJS..220...15P} {220, 15}

\bibitem[\protect\citeauthoryear{{Paxton} et~al.,}{{Paxton}
  et~al.}{2018}]{Paxton2018}
{Paxton} B.,  et~al., 2018, \mn@doi [\apjs] {10.3847/1538-4365/aaa5a8}, \href
  {https://ui.adsabs.harvard.edu/abs/2018ApJS..234...34P} {234, 34}

\bibitem[\protect\citeauthoryear{{Paxton} et~al.,}{{Paxton}
  et~al.}{2019}]{Paxton2019}
{Paxton} B.,  et~al., 2019, \mn@doi [\apjs] {10.3847/1538-4365/ab2241}, \href
  {https://ui.adsabs.harvard.edu/abs/2019ApJS..243...10P} {243, 10}

\bibitem[\protect\citeauthoryear{{Pejcha}, {Metzger}  \& {Tomida}}{{Pejcha}
  et~al.}{2016a}]{2016MNRAS.455.4351P}
{Pejcha} O.,  {Metzger} B.~D.,   {Tomida} K.,  2016a, \mn@doi [\mnras]
  {10.1093/mnras/stv2592}, \href
  {https://ui.adsabs.harvard.edu/abs/2016MNRAS.455.4351P} {455, 4351}

\bibitem[\protect\citeauthoryear{{Pejcha}, {Metzger}  \& {Tomida}}{{Pejcha}
  et~al.}{2016b}]{2016MNRAS.461.2527P}
{Pejcha} O.,  {Metzger} B.~D.,   {Tomida} K.,  2016b, \mn@doi [\mnras]
  {10.1093/mnras/stw1481}, \href
  {https://ui.adsabs.harvard.edu/abs/2016MNRAS.461.2527P} {461, 2527}

\bibitem[\protect\citeauthoryear{{Pejcha}, {Metzger}, {Tyles}  \&
  {Tomida}}{{Pejcha} et~al.}{2017}]{2017ApJ...850...59P}
{Pejcha} O.,  {Metzger} B.~D.,  {Tyles} J.~G.,   {Tomida} K.,  2017, \mn@doi
  [\apj] {10.3847/1538-4357/aa95b9}, \href
  {https://ui.adsabs.harvard.edu/abs/2017ApJ...850...59P} {850, 59}

\bibitem[\protect\citeauthoryear{{Pelisoli}, {Vos}, {Geier}, {Schaffenroth}  \&
  {Baran}}{{Pelisoli} et~al.}{2020}]{2020A&A...642A.180P}
{Pelisoli} I.,  {Vos} J.,  {Geier} S.,  {Schaffenroth} V.,   {Baran} A.~S.,
  2020, \mn@doi [\aap] {10.1051/0004-6361/202038473}, \href
  {https://ui.adsabs.harvard.edu/abs/2020A&A...642A.180P} {642, A180}

\bibitem[\protect\citeauthoryear{{Pols}, {Tout}, {Eggleton}  \& {Han}}{{Pols}
  et~al.}{1995}]{Pols1995}
{Pols} O.~R.,  {Tout} C.~A.,  {Eggleton} P.~P.,   {Han} Z.,  1995, \mn@doi
  [\mnras] {10.1093/mnras/274.3.964}, \href
  {https://ui.adsabs.harvard.edu/abs/1995MNRAS.274..964P} {274, 964}

\bibitem[\protect\citeauthoryear{{Potekhin} \& {Chabrier}}{{Potekhin} \&
  {Chabrier}}{2010}]{Potekhin2010}
{Potekhin} A.~Y.,  {Chabrier} G.,  2010, \mn@doi [Contributions to Plasma
  Physics] {10.1002/ctpp.201010017}, \href
  {https://ui.adsabs.harvard.edu/abs/2010CoPP...50...82P} {50, 82}

\bibitem[\protect\citeauthoryear{{Price} \& {Monaghan}}{{Price} \&
  {Monaghan}}{2007}]{2007MNRAS.374.1347P}
{Price} D.~J.,  {Monaghan} J.~J.,  2007, \mn@doi [\mnras]
  {10.1111/j.1365-2966.2006.11241.x}, \href
  {https://ui.adsabs.harvard.edu/abs/2007MNRAS.374.1347P} {374, 1347}

\bibitem[\protect\citeauthoryear{{Reichardt}, {De Marco}, {Iaconi}, {Tout}  \&
  {Price}}{{Reichardt} et~al.}{2019}]{2019MNRAS.484..631R}
{Reichardt} T.~A.,  {De Marco} O.,  {Iaconi} R.,  {Tout} C.~A.,   {Price}
  D.~J.,  2019, \mn@doi [\mnras] {10.1093/mnras/sty3485}, \href
  {https://ui.adsabs.harvard.edu/abs/2019MNRAS.484..631R} {484, 631}

\bibitem[\protect\citeauthoryear{{Rogers} \& {Nayfonov}}{{Rogers} \&
  {Nayfonov}}{2002}]{Rogers2002}
{Rogers} F.~J.,  {Nayfonov} A.,  2002, \mn@doi [\apj] {10.1086/341894}, \href
  {https://ui.adsabs.harvard.edu/abs/2002ApJ...576.1064R} {576, 1064}

\bibitem[\protect\citeauthoryear{{Saumon}, {Chabrier}  \& {van Horn}}{{Saumon}
  et~al.}{1995}]{Saumon1995}
{Saumon} D.,  {Chabrier} G.,   {van Horn} H.~M.,  1995, \mn@doi [\apjs]
  {10.1086/192204}, \href
  {https://ui.adsabs.harvard.edu/abs/1995ApJS...99..713S} {99, 713}

\bibitem[\protect\citeauthoryear{{Schuessler} \& {Schmitt}}{{Schuessler} \&
  {Schmitt}}{1981}]{1981A&A....97..373S}
{Schuessler} I.,  {Schmitt} D.,  1981, \aap, \href
  {https://ui.adsabs.harvard.edu/abs/1981A&A....97..373S} {97, 373}

\bibitem[\protect\citeauthoryear{{Semenov}, {Henning}, {Helling}, {Ilgner}  \&
  {Sedlmayr}}{{Semenov} et~al.}{2003}]{2003A&A...410..611S}
{Semenov} D.,  {Henning} T.,  {Helling} C.,  {Ilgner} M.,   {Sedlmayr} E.,
  2003, \mn@doi [\aap] {10.1051/0004-6361:20031279}, \href
  {https://ui.adsabs.harvard.edu/abs/2003A&A...410..611S} {410, 611}

\bibitem[\protect\citeauthoryear{{Smarr} \& {Blandford}}{{Smarr} \&
  {Blandford}}{1976}]{1976ApJ...207..574S}
{Smarr} L.~L.,  {Blandford} R.,  1976, \mn@doi [\apj] {10.1086/154524}, \href
  {https://ui.adsabs.harvard.edu/abs/1976ApJ...207..574S} {207, 574}

\bibitem[\protect\citeauthoryear{{Soker} \& {Kashi}}{{Soker} \&
  {Kashi}}{2012}]{2012ApJ...746..100S}
{Soker} N.,  {Kashi} A.,  2012, \mn@doi [\apj] {10.1088/0004-637X/746/1/100},
  \href {https://ui.adsabs.harvard.edu/abs/2012ApJ...746..100S} {746, 100}

\bibitem[\protect\citeauthoryear{{Soker} \& {Rappaport}}{{Soker} \&
  {Rappaport}}{2001}]{2001ApJ...557..256S}
{Soker} N.,  {Rappaport} S.,  2001, \mn@doi [\apj] {10.1086/321669}, \href
  {https://ui.adsabs.harvard.edu/abs/2001ApJ...557..256S} {557, 256}

\bibitem[\protect\citeauthoryear{{Soker} \& {Tylenda}}{{Soker} \&
  {Tylenda}}{2003}]{2003ApJ...582L.105S}
{Soker} N.,  {Tylenda} R.,  2003, \mn@doi [\apjl] {10.1086/367759}, \href
  {https://ui.adsabs.harvard.edu/abs/2003ApJ...582L.105S} {582, L105}

\bibitem[\protect\citeauthoryear{Springel et~al.,}{Springel
  et~al.}{2005}]{Springel_2005}
Springel V.,  et~al., 2005, \mn@doi [Nature] {10.1038/nature03597}, 435,
  629–636

\bibitem[\protect\citeauthoryear{{St{\c e}pie{\'n}}}{{St{\c
  e}pie{\'n}}}{2011}]{2011A&A...531A..18S}
{St{\c e}pie{\'n}} K.,  2011, \mn@doi [\aap] {10.1051/0004-6361/201116689},
  \href {http://adsabs.harvard.edu/abs/2011A%26A...531A..18S} {531, A18}

\bibitem[\protect\citeauthoryear{{Stevenson}, {Vigna-G{\'o}mez}, {Mandel},
  {Barrett}, {Neijssel}, {Perkins}  \& {de Mink}}{{Stevenson}
  et~al.}{2017}]{2017NatCo...814906S}
{Stevenson} S.,  {Vigna-G{\'o}mez} A.,  {Mandel} I.,  {Barrett} J.~W.,
  {Neijssel} C.~J.,  {Perkins} D.,   {de Mink} S.~E.,  2017, \mn@doi [Nature
  Communications] {10.1038/ncomms14906}, \href
  {https://ui.adsabs.harvard.edu/abs/2017NatCo...814906S} {8, 14906}

\bibitem[\protect\citeauthoryear{{Timmes} \& {Swesty}}{{Timmes} \&
  {Swesty}}{2000}]{Timmes2000}
{Timmes} F.~X.,  {Swesty} F.~D.,  2000, \mn@doi [\apjs] {10.1086/313304}, \href
  {https://ui.adsabs.harvard.edu/abs/2000ApJS..126..501T} {126, 501}

\bibitem[\protect\citeauthoryear{{Tutukov} \& {Yungelson}}{{Tutukov} \&
  {Yungelson}}{1993}]{1993MNRAS.260..675T}
{Tutukov} A.~V.,  {Yungelson} L.~R.,  1993, \mn@doi [\mnras]
  {10.1093/mnras/260.3.675}, \href
  {https://ui.adsabs.harvard.edu/abs/1993MNRAS.260..675T} {260, 675}

\bibitem[\protect\citeauthoryear{{Tylenda} et~al.,}{{Tylenda}
  et~al.}{2011}]{2011A&A...528A.114T}
{Tylenda} R.,  et~al., 2011, \mn@doi [\aap] {10.1051/0004-6361/201016221},
  \href {http://adsabs.harvard.edu/abs/2011A&A...528A.114T} {528, A114}

\bibitem[\protect\citeauthoryear{{Vigna-G{\'o}mez} et~al.,}{{Vigna-G{\'o}mez}
  et~al.}{2018}]{2018MNRAS.481.4009V}
{Vigna-G{\'o}mez} A.,  et~al., 2018, \mn@doi [\mnras] {10.1093/mnras/sty2463},
  \href {https://ui.adsabs.harvard.edu/abs/2018MNRAS.481.4009V} {481, 4009}

\bibitem[\protect\citeauthoryear{{Voss} \& {Tauris}}{{Voss} \&
  {Tauris}}{2003}]{2003MNRAS.342.1169V}
{Voss} R.,  {Tauris} T.~M.,  2003, \mn@doi [\mnras]
  {10.1046/j.1365-8711.2003.06616.x}, \href
  {https://ui.adsabs.harvard.edu/abs/2003MNRAS.342.1169V} {342, 1169}

\bibitem[\protect\citeauthoryear{{van den Heuvel}}{{van den
  Heuvel}}{1976}]{1976IAUS...73...35V}
{van den Heuvel} E.~P.~J.,  1976, in {Eggleton} P.,  {Mitton} S.,   {Whelan}
  J.,  eds,  IAU Symposium Vol. 73, Structure and Evolution of Close Binary
  Systems. p.~35

\makeatother
\end{thebibliography}
